\documentclass[reprint,showpacs,showkeys,superscriptaddress,aps,pra]{revtex4-1}

\usepackage{graphicx}
\usepackage{epstopdf}
\usepackage{bm}
\usepackage{amssymb}
\usepackage{amsmath}
\usepackage{bbold}
\usepackage{ulem}
\usepackage{color, soul}
\usepackage{xcolor}
\usepackage{afterpage}
\usepackage{footmisc}
\usepackage{natbib}
\usepackage{float}
\usepackage{multirow}

\begin{document}
	
	\preprint{APS/123-QED}
	
	\title{An ac-susceptibility study of the magnetic relaxation phenomena in antiskyrmion host tetragonal Mn-Pt(Pd)-Sn system }
	\author{P. V. Prakash Madduri}%
	\affiliation{School of Physical Sciences, National Institute of Science Education and Research, HBNI, Jatni-752050, India}
	\author{Subir Sen}%
	\affiliation{School of Physical Sciences, National Institute of Science Education and Research, HBNI, Jatni-752050, India}
	\author{Bimalesh Giri}%
	\affiliation{School of Physical Sciences, National Institute of Science Education and Research, HBNI, Jatni-752050, India}
	\author{Dola Chakrabartty}%
	\affiliation{School of Physical Sciences, National Institute of Science Education and Research, HBNI, Jatni-752050, India}
	\author{Subhendu K. Manna}%
	\affiliation{School of Physical Sciences, National Institute of Science Education and Research, HBNI, Jatni-752050, India}
	\author{Stuart S. P. Parkin}
	\affiliation{Max Planck Institute of Microstructure Physics, Weinberg 2, 06120 Halle, Germany}
	\author{Ajaya K. Nayak}
	\email{ajaya@niser.ac.in}
	\affiliation{School of Physical Sciences, National Institute of Science Education and Research, HBNI, Jatni-752050, India}

	\date{\today}% It is always \today, today,

\begin{abstract}
 Here, we report an exhaustive study of the frequency-dependent ac-magnetic susceptibility of the $D_{2d}$ symmetric Heusler system Mn-Pt(Pd)-Sn that hosts antiskyrmions over a wide temperature range. Magnetic relaxation studies using Cole-Cole formalism reveal a Debye-type relaxation with a nearly negligible distribution in relaxation times. In contrast to the archetypical skyrmion hosts, the high Curie temperature ($ T_C $) of the present system ensures shorter switching times, and, correspondingly, higher frequencies are required to probe the relaxation dynamics. We find a non-monotonic variation in the characteristic relaxation time with distinct maxima at the phase boundaries \textit{via} helical $\longrightarrow$ antiskyrmion $\longrightarrow$ field-polarized states, indicating slower magnetization dynamics over the region of phase coexistence. The temperature-dependent relaxation time across different phases is of the order of $ 10^{-5} - 10^{-4} $ s and follows the well-known Arrhenius law with reasonable values of the energy barriers. The present study concerning the magnetization dynamics in the antiskyrmion host tetragonal Heusler system is an important contribution towards the basic understanding of the dynamical aspects of antiskyrmions for their potential applications.
\end{abstract}

\maketitle

%\tableofcontents

\section{INTRODUCTION}
Skyrmions represent a particular class of magnetic \textit{nano-domains} characterized by vortex-like localized spin configurations \cite{U.K.Robler}, in contrast to the conventional- helical, conical, and spontaneously magnetized spin structures. The twist in the magnetization profile within each domain described by a topological winding number gives robust protection to the spin configurations against continuous deformation \cite{N.Nagaosa}. Owing to their solitonic nature, skyrmions can interact efficiently with the electrons and magnons in the host material and result in several exotic electromagnetic phenomena such as topological Hall effect \cite{A.Neubauer}, thermally induced ratchet motion \cite{M.Mochizuki}, and skyrmion magnetic resonance \cite{Y.Onose}. The competition between the Heisenberg exchange and the Dzyaloshinskii-Moriya (DM) interaction is one of the main reasons behind the observation of skyrmions in bulk and thin-film magnetic systems \cite{A.Neubauer,S.Muhlbauer,C.Pappas,W.Munzer,X.Z.Yu,S.X.Huang,H.Wilhelm,S.Seki,S.Heinze}. Recently, a new type of topological object named antiskyrmion was observed in an acentric tetragonal Heusler-Mn-Pt(Pd)-Sn system with $D_{2d}$ crystal symmetry \cite{A.K.Nayak}. Note that the high $ T_C $ of this material ($ \approx $ 400 K) ensures that the  antiskyrmions get stabilized over a wide temperature range. In the archetypal skyrmion hosting   B20 materials, an externally applied magnetic field favours conical helices propagating along the field direction. In contrast, a particular pattern of DM vectors dictated by the $ D_{2d} $ crystallographic symmetry in the Mn-Pt(Pd)-Sn system precludes the emergence of the longitudinal conical structure at finite fields.\\  

In most cases, the presence of the skyrmion phase was identified using small-angle neutron scattering \cite{S.Muhlbauer}, Lorentz transmission electron microscopy (LTEM) \cite{X.Z.Yu, A.K.Nayak} and, indirectly  by topological Hall transport studies \cite{A.Neubauer,N.Kanazava,S.X.Huang,S.X.Huang2,S.Sen}. It is well-known that ac-susceptibility is a potential technique to identify various magnetic phases and to study their relaxation dynamics \cite{A.H.Morrish,M.Balanda,S.J.Blundell}. Linear and non-linear susceptibilities with various ac-magnetic field amplitudes ($ H_{ac} $) and frequencies ($ f $) in the presence and absence of superimposed dc-magnetic fields ($ H_{dc} $) are widely used to unambiguously identify Ferro-, Ferri-, Antiferro-magnetic ordering, canonical-, cluster-, reentrant spin glass nature, spin-reorientation transitions and superparamagnetism  \cite{M.Balanda,S.J.Blundell}. Recently, it has been shown that ac-susceptibility as a function of dc-magnetic field, $ \chi\left( H\right)  $, at various fixed frequencies (\textit{f}) of the oscillating magnetic field can be used to study the relaxation dynamics of the modulated magnetic phases in different skyrmion host materials that include metallic- FeGe \cite{H.Wilhelm}, MnSi \cite{A.Bauer}, Fe$ _{1-x} $Co$ _{x} $Si \cite{L.J.Bannenberg}, Mn$ _{1-x} $Fe$ _{x} $Si \cite{L.J.Bannenberg2}, insulating Cu$ _{2} $OSeO$ _{3} $\cite{I.Levatic,F.Qian} and semiconducting GaV$ _{4} $S$_{8}$\cite{A.Butykai}. Peak/hump anomalies in the field evolution of ac-susceptibility that characterize the antiskyrmion phase in the tetragonal Heusler Mn-Pt(Pd)-Sn were recently reported \cite{S.K.Jamal}. However, detailed studies are required to understand the magnetic relaxation and magnetic phase transitions in such a system. To this end, we have undertaken an in-depth study of high-precision ac-susceptibility data on the antiskyrmion hosts-Mn$_{1.4}$PtSn and  Mn$_{1.4}$Pt$_{0.9}$Pd$_{0.1}$Sn.

%%%%%%%%%%%%%%%%%%%%%%%%%%%%%%%%%%%%%%%%%%%%%%%%%%%%%%%%%%%%%%%%%%%%%%%%%%%%%%%%%%%%%%%%%%%%%%%%%%%%%%%%%%%%%%%%%%%%%%%%%%%%%%%%%%%%%%%%%%%%%%%%%%%%%%

	\begin{figure} [tb!]
	\includegraphics[angle=0,width=9cm,clip]{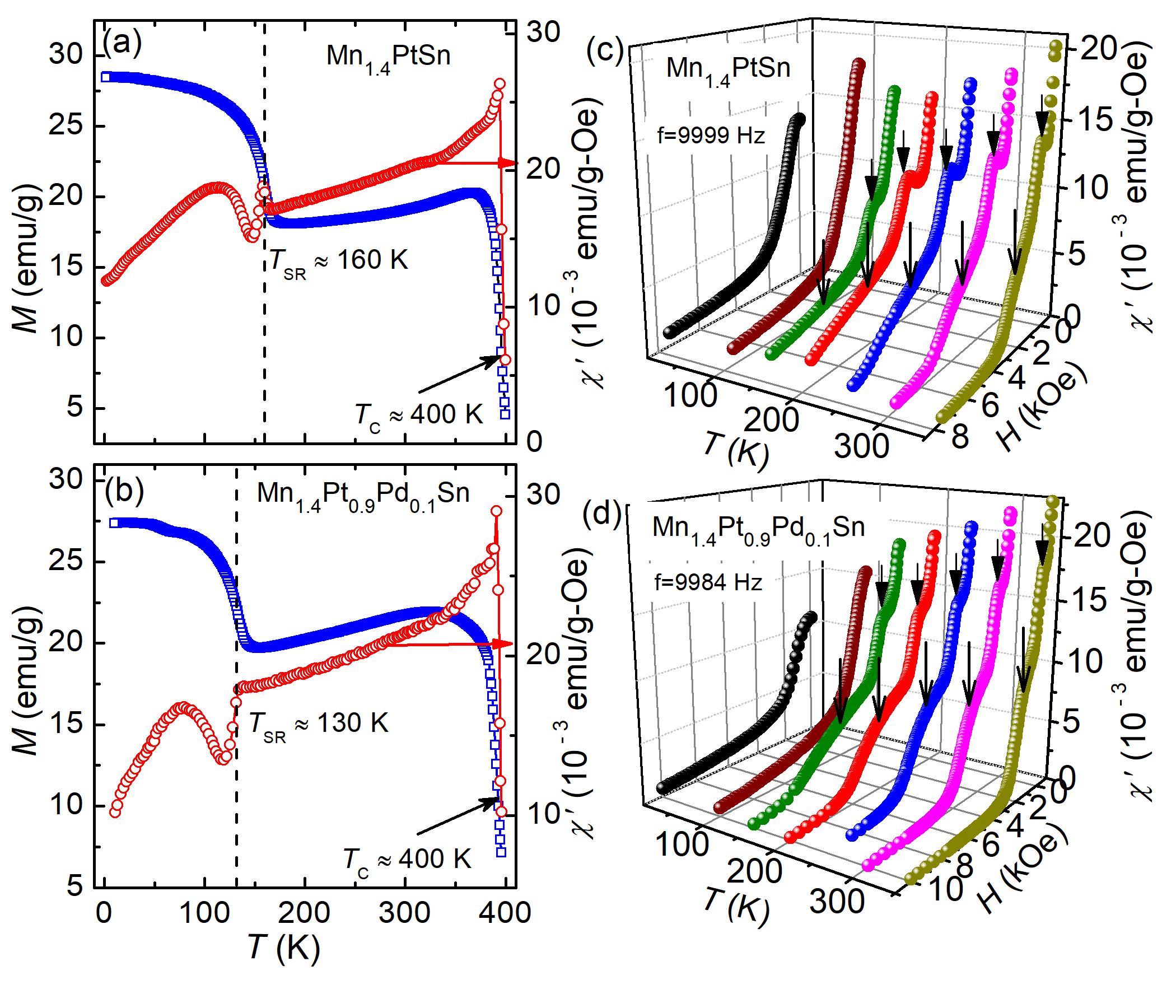}
	\caption{\label{FIG1}(Color online) Magnetization  as a function of temperature, $ M(T) $, (open squares, left Y-axis) measured at $ H $ = 1 kOe in field-cooled mode and the temperature dependent real part of zero-field ac-susceptibility curves $ \chi^{\prime}(T) $ shown with open circles (right Y-axis) for (a) Mn$_{1.4}$PtSn and (b) Mn$_{1.4}$Pt$_{0.9}$Pd$_{0.1}$Sn. Isothermal field dependent ac-susceptibility curves $ \chi(H) $ taken at different temperatures in the range 10 K $ \leq T \leq $ 350 K for (c) Mn$_{1.4}$PtSn and (d) Mn$_{1.4}$Pt$_{0.9}$Pd$_{0.1}$Sn. The evolution of antiskyrmion phase is inferred from the broad minimum  surrounded by maxima in $ \chi(H) $ curves. The low field maxima is well seen by the occurrence of a peak, whereas, the high field maxima is visualized by a speared out hump kind of behavior marked by arrows.}
	\label{Fig1}
\end{figure}

%%%%%%%%%%%%%%%%%%%%%%%%%%%%%%%%%%%%%%%%%%%%%%%%%%%%%%%%%%%%%%%%%%%%%%%%%%%%%%%%%%%%%%%%%%%%%%%%%%%%%%%%%%%%%%%%%%%%%%%%%%%%%%%%%%%%%%%%%%%%%%%%%%%%%%
\section{Methods}
Polycrystalline ingots of Mn$_{1.4}$PtSn and Mn$_{1.4}$Pt$_{0.9}$Pd$_{0.1}$Sn were synthesized using arc melting technique. Details of the sample preparation and structural characterization are provided in the supplementary information \cite{Supp}. Nearly disk-shaped samples of Mn$_{1.4}$PtSn and Mn$_{1.4}$Pt$_{0.9}$Pd$_{0.1}$Sn with a mass of 80.30 mg and 100.2 mg, respectively, were taken for the ac-susceptibility measurements performed on a Quantum Design Physical Property Measurement System. At fixed temperatures (\textit{T}), ac-susceptibility ($ \chi $) as a function of the magnetic field (\textit{H})  was measured at an r.m.s field ($ H_{ac} $) of 10 Oe and frequencies ($ f $) ranging from 11 Hz to 9999 Hz.  For each measurement, the sample was `zero-field-cooled' from 400 K ($ \approx T_{C} $) to the desired temperature at a rate of 10~K/min. The in-phase and out-of-phase components of the oscillating magnetization were measured and normalized by the ac drive amplitude ($ H_{ac} $) to obtain the real ($ \chi^{\prime} $) and imaginary ($ \chi^{\prime\prime}$) parts of the ac-susceptibility, respectively. Note that during the $\chi(H)$ scan, the magnetic field was stabilized at each step in the \textit{linear approach} mode to achieve high precision data. 

%%%%%%%%%%%%%%%%%%%%%%%%%%%%%%%%%%%%%%%%%%%%%%%%%%%%%%%%%%%%%%%%%%%%%%%%%%%%%%%%%%%%%%%%%%%%%%%%%%%%%%%%%%%%%%%%%%%%%%%%%%%%%%%%%%%%%%%%%%%%%%%%%%%%%%
\section{RESULTS AND DISCUSSION}
%%%%%%%%%%%%%%%%%%%%%%%%%%%%%%%%%%%%%%%

Mn$_{1.4}$PtSn exhibits a $T_{C}$ of about 400 K and  a low-temperature spin-reorientation transition [$ T_{SR}$] of $\approx $ 160 K, as depicted in the temperature dependent magnetization $ M(T) $ data plotted in Fig. \ref{Fig1}(a). The onset of $ T_{SR} $ brings about a sudden change in the magnetization in the $ M(T) $ data and  a peak/dip kind of behavior in the temperature-dependent real part of ac-susceptibility $ \chi^{\prime}(T) $ curve [Fig. 1(a)]. In case of Mn$_{1.4}$Pt$_{0.9}$Pd$_{0.1}$Sn, the $T_{C}$ is found at a temperature of about 400~K and  the $ T_{SR} $ at 130~K [Fig. \ref{Fig1}(b)]. It has been reported that a change in the magnetic structure below the $ T_{SR} $  hinders the nucleation process of the antiskyrmion phase \cite{A.K.Nayak}. Similar inference has also been made from the field-dependent ac-susceptibility and magnetic entropy change measurements in the same materials \cite{S.K.Jamal}. Since the antiskyrmion phase in the present system is well-established between the $ T_{SR} $ and the $T_{C}$, the present work mostly focuses on the ac-susceptibility analysis in this particular temperature range. Figure \ref{Fig1}(c) and (d)  show the field dependent ac-susceptibility  [$ \chi(H) $] data measured at different temperatures starting from 10 K to 350 K. The presence of noticeable peak/dip kind of anomalies in the $ \chi(H) $ curves suggest that the applied dc-magnetic fields gradually transform the system into different magnetically ordered states. As it can be seen,  the $ \chi(H) $ curve initially starts decreasing with increasing field before exhibiting a maxima  in the field range of 1 kOe - 1.5 kOe for Mn$_{1.4}$PtSn and 0.6 kOe - 1.3 kOe for Mn$_{1.4}$Pt$_{0.9}$Pd$_{0.1}$Sn (marked by arrows in the low-field regime). Further increasing the field gives rise to  a broad hump like behavior (marked by arrows in the intermidiate-field regime) in the $ \chi(H) $ data before  decreasing monotonically at higher fields. This broad hump in the $ \chi(H) $ data is well pronounced at temperatures close to the $ T_{C} $.  The corresponding peak/hump positions  signify lower ($ H_{AskX}^{L} $) and upper ($ H_{AskX}^{H} $) critical fields that bound the antiskyrmion (AskX) phase. For $ H < H_{AskX}^{L} $, the magnetic state is characterized by a helical spin modulation (H), while for $H > H_{AskX}^{H} $, it is a collinear/field-polarized (FP) state. It can be noted here that no peak/hump kind of behavior is found for the $ \chi(H) $ curves measured at 100~K and 2~K, signifying the absence of antiskyrmion phase in this temperature regime.

%%%%%%%%%%%%%%%%%%%%%%%%%%%%%%%%%%%%%%%%%%%%%%%%%%%%%%%%%%%%%%%%%%%%%%%%%%%%%%%%%%%%%%%%%%%%%%%%%%%%%%%%%%%%%%%%%%%%%%%%%%%%%%%%%%%%%%%%%%%%%%%%%%%%%%
	\begin{figure*} [tb!] 
	\includegraphics[angle=0,width=14cm,clip]{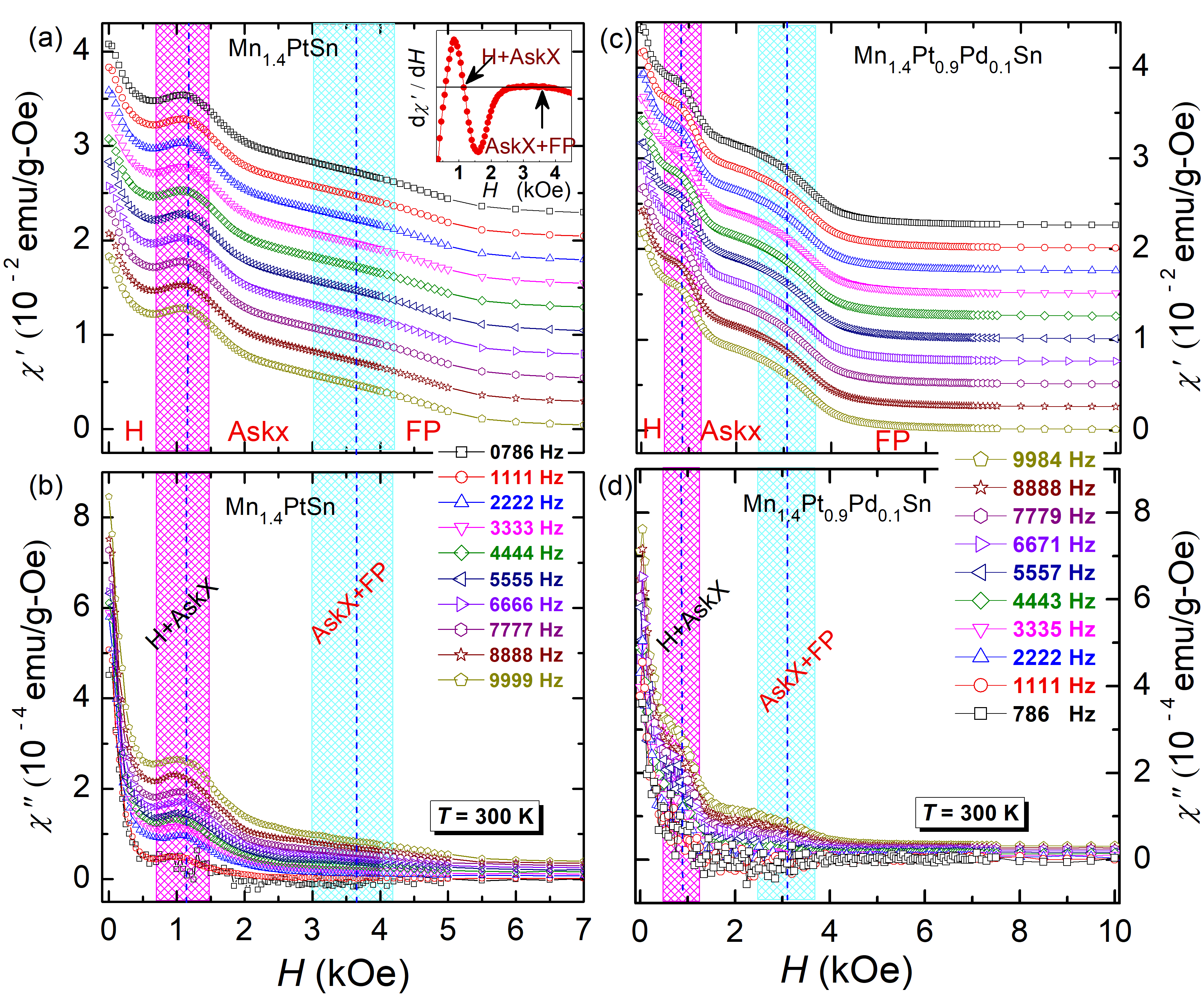}
	\caption{\label{FIG2}(Color online)  In-phase [$ \chi^{\prime}(H) $] and out-of-phase [$ \chi^{\prime\prime}(H) $] components of $ \chi(H) $ at $ T $ = 300 K with $ H_{ac}$ = 10 Oe measured at different frequencies for Mn$_{1.4}$PtSn [panels (a) and (b)] and  Mn$_{1.4}$Pt$_{0.9}$Pd$_{0.1}$Sn [panels (c) and (d)]. Note that for the sake of clarity, the successive $ \chi^{\prime}(H) $ data starting from 8888 Hz, have been given a constant upward shift of 2.5 m emu/g-Oe. Vertical dashed lines represent phase crossover from H-AskX and AskX-FP states inferred from the d$\chi^{\prime}$/d$H$ \textit{versus} \textit{H} as shown as example in the inset of (a). Shaded regions highlights the co-existence of multiple phases.}
	\label{Fig2}
\end{figure*}
%%%%%%%%%%%%%%%%%%%%%%%%%%%%%%%%%%%%%%%%%%%%%%%%%%%%%%%%%%%%%%%%%%%%%%%%%%%%%%%%%%%%%%%%%%%%%%%%%%%%%%%%%%%%%%%%%%%%%%%%%%%%%%%%%%%%%%%%%%%%%%%%%%%%%%

  The field dependence of  $ \chi^{\prime}(H) $ and $ \chi^{\prime\prime}(H) $ data measured at 300 K in presence of different frequencies are plotted in Fig. \ref{Fig2}. As highlighted by the shaded regions,   both $ \chi^{\prime} $ and $ \chi^{\prime\prime} $  exhibit hump kind of behaviour at the phase transition. Note that the magnitude of $ \chi^{\prime}(H) $ is almost independent of the frequency (\textit{f}) of the $ H_{ac} $, especially when $f \lesssim $ 500 Hz (see Fig. S3) \cite{Supp}. At low frequencies the $ \chi^{\prime\prime}(H) $ response is so weak that the data are statistically scattered and no information can be gained. It is also found that the $ \chi^{\prime\prime}(H) $ signal is of about two orders of magnitude smaller than that of $ \chi^{\prime}(H) $ and falls close to the instrument resolution limit. 
  
%%%%%%%%%%%%%%%%%%%%%%%%%%%%%%%%%%%%%%%%%%%%%%%%%%%%%%%%%%%%%%%%%%%%%%%%%%%%%%%%%%%%%%%%%%%%%%%%%%%%%%%%%%%%%%%%%%%%%%%%%%%%%%%%%%%%%%%%%%%%%%%%%%%%

%%%%%%%%%%%%%%%%%%%%%%%%%%%%%%%%%%%%%%%%%%%%%%%%%%%%%%%%%%%%%%%%%%%%%%%%%%%%%%%%%%%%%%%%%%%%%%%%%%%%%%%%%%%%%%%%%%%%%%%%%%%%%%%%%%%%%%%%%%%%%%%%%%%%%
\begin{figure*} [tb!]
	\includegraphics[angle=0,width=17cm,clip]{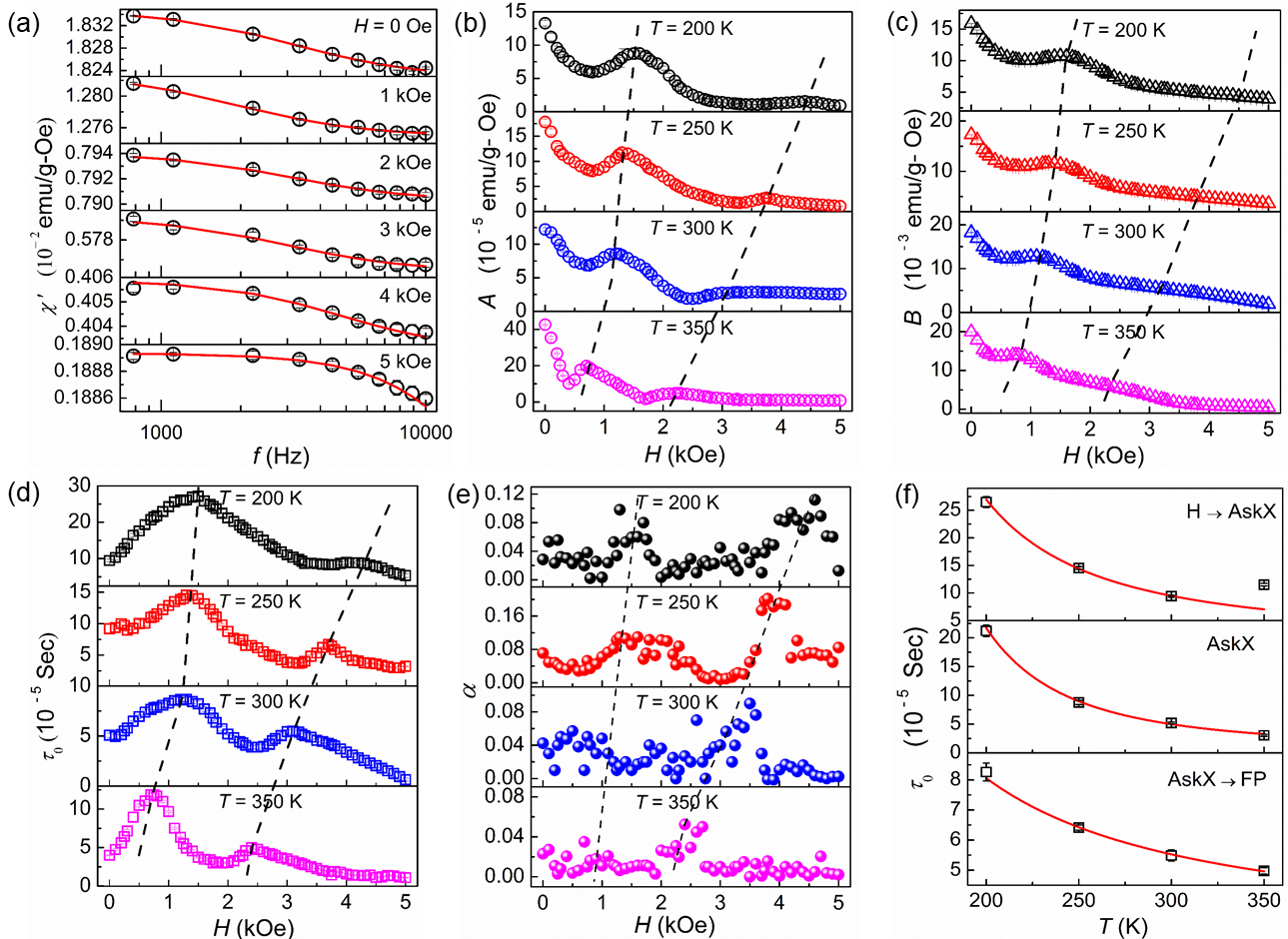}
	\caption{\label{FIG3}(Color online)  (a) Frequency dependence of $ \chi^{\prime} $ at different representative magnetic fields ($ H $) measured at $ T $ = 300 K for Mn$ _{1.4} $PtSn. The fits (solid lines) to the data (open symbols) are based on the Eq. (\ref{Eq.1}) described in the text. Magnetic field dependence of the fitting parameters for Mn$ _{1.4} $PtSn (b) $ \chi_{T} - \chi_{S} $, (c) $B = \chi_{S} $, (d) the characteristic relaxation time ($ \tau_{0} $), and (e) the relaxation times distribution parameter, $ \alpha $. Vertical dashed lines serve as guide to eye to indicate the phase boundaries between H, AskX and FP phases. (f) Temperature dependence of relaxation time $ \tau_{0} $ derived from (d). $ \tau_{0} $ at H $\rightarrow$ AskX and AskX $\rightarrow$ FP regions are inferred from the lower and upper maxima in the $ \tau_{0}(H) $ plots, whereas the $ \tau_{0}(T)$ in the AskX phase is extracted from $ \tau_{0}(H) $ at \textit{H} = 2 kOe. Note that the continuous lines represent fits to the observed data (open sqaure symbols) using Arrhenius law.}
	\label{Fig3}
\end{figure*}
%%%%%%%%%%%%%%%%%%%%%%%%%%%%%%%%%%%%%%%%%%%%%%%%%%%%%%%%%%%%%%%%%%%%%%%%%%%%%%%%%%%%%%%%%%%%%%%%%%%%%%%%%%%%%%%%%%%%%%%%%%%%%%%%%%%%%%%%%%%%%%%%%%%%

The $ \chi^{\prime}(H) $ data at different frequencies over 786 Hz $ \leqslant ~f~ \leqslant$ 9999 Hz is re-plotted as $ \chi^{\prime}(f) $ to systematically investigate the frequency dependency of $ \chi^{\prime}$ at various static magnetic fields covering the helical, AskX and field-polarized magnetic phases. First, we analyze the data for Mn$ _{1.4} $PtSn as shown in Fig. \ref{Fig3}(a). The important observations are as follows:

(i) An inflection point in the $ \chi^{\prime}(f) $ data indicates that the relaxation in the antiskyrmion host-Mn$ _{1.4} $PtSn happens in a frequency window covered in the present experiment. 

(ii) In sharp contrast to the skyrmion hosts Fe$_{1-x}$Co$_{x}$Si \cite{L.J.Bannenberg}, Cu$_{2}$OSeO$_{3}$ \cite{F.Qian}, and GaV$_{4}$S$_{8}$ \cite{A.Butykai}, where the entire magnetic relaxation is observed in the frequency range over 0.1 Hz to 1 kHz, a frequency window ranging from 0.5 kHz to 10 kHz is needed to study the magnetization dynamics in the  Mn$ _{1.4} $PtSn. A possible reason for the observed behavior could be comparably high  $T_{C}$ ($\approx$ 400 K) that signifies a large exchange energy \textit{J} in the present system. According to Heisenberg's principle, higher exchange energies leads to shorter switching times as $ \tau ~\propto ~\hslash/J $. Hence higher frequencies are needed in order to probe the relaxation dynamics of the present system.

(iii) Noticeable changes are observed in the $ \chi^{\prime}\left(f\right)$ as one moves gradually from lower to higher magnetic fields. At any given temperature, the inflection point in the  $ \chi^{\prime}(f) $ shifts to higher frequencies with increasing fields [Fig. \ref{Fig3}(a)]. 

%%%%%%%%%%%%%%%%%%%%%%%%%%%%%%%%%%%%%%%%%%%%%%%%%%%%%%%%%%%%%%%%%%%%%%%%%%%%%%%%%%%%%%%%%%%%%%%%%%%%%%%%%%%%%%%%%%%%%%%%%%%%%%%%%%%%%%%%%%%%%%%%%%%%

We have attempted to fit the observed variation of $ \chi^{\prime}\left(f\right)$ and  $\chi^{\prime\prime}\left(f\right)$ data using the well-known Cole-Cole relation \cite{P.Debye,Casimir,Cole.Cole} as given below,
\begin{equation}
 \chi^{\prime}(f) =  \chi_{S}+\frac{A~\left[1+\left(2\pi f\tau_{0}\right)^{1-\alpha} \sin\left( \dfrac{\pi\alpha}{2}\right) \right] }{1+2\left(2\pi f\tau_{0}\right)^{1-\alpha} \sin \left( \dfrac{\pi\alpha}{2}\right)+\left(2\pi f\tau_{0}\right)^{2\left( {1-\alpha} \right)} }\label{Eq.1}
\end{equation}
\begin{equation}
\chi^{\prime\prime}(f) =\frac{A~\left[\left(2\pi f\tau_{0}\right)^{1-\alpha} \cos\left( \dfrac{\pi\alpha}{2}\right) \right] }{1+2\left(2\pi f\tau_{0}\right)^{1-\alpha} \sin \left( \dfrac{\pi\alpha}{2}\right)+\left(2\pi f\tau_{0}\right)^{2\left( {1-\alpha} \right)} }\label{Eq.2}
\end{equation}
where \textit{A} = $ \chi_{T} - \chi_{S} $ with $ \chi_{S} \equiv \chi_{f\longrightarrow \infty}$, $\chi_{T} \equiv \chi_{f\longrightarrow 0} $ represent adiabatic and isothermal susceptibilities, $ \tau_{0} $ is the average relaxation time, and `$ \alpha $' characterizes the width of the distribution in the relaxation times.

 It can be clearly seen from Fig. \ref{Fig3}(a) that the fits based on Eq. (\ref{Eq.1}) best represent the observed $ \chi^{\prime}\left(f\right)$ data. The fit parameters $ A $, $ B \equiv \chi_{S}$, $ \tau_{0} $, and $ \alpha $ are obtained at each field step. The magnetic field variation of the fit parameters at different temperatures are shown in Fig. \ref{Fig3} (b)-(e). The following conclusions can be inferred from the field variation of the fit parameters based on the Cole-Cole formalism. 

(i) A non-monotonic variation is observed in the  field  dependency of $ A  $, $ B  (\equiv \chi_{S})$, $ \tau_{0} $, and $ \alpha $. A broad hump kind of behaviour in the field variations of the fit parameters at magnetic field \textit{H} = $H_{1}$ [left dashed curves in Fig. \ref{Fig3}(b)-(e)] and $H_{2}$ [right dashed curves in Fig. \ref{Fig3}(b)-(e)] is found. These $H_{1}$ and $H_{2}$ are temperature sensitive and show a good agreement with the $ H_{AskX}^{L}$ and $ H_{AskX}^{H} $ inferred from the $ \chi^{\prime}(H)$, $ \chi^{\prime\prime}(H)$. 

(ii) From figure \ref{Fig3}(b), the parameter $ \chi_{T} - \chi_{S} $ shows broad maxima  at \textit{H} = $H_{1}$ and $H_{2}$. Higher values of `$ A $' highlight prominent magnetic relaxation in the region of the phase coexistence (helical-to-AskX and AskX-to-FP). Butykai \textit{et al.} \cite{A.Butykai} attributed this kind of behaviour to the defects in the magnetization profile arising from the incommensurate spin modulations on the verge of magnetic phase transformation. 
%It was reported that $ \chi_{T} - \chi_{S} $ $\approx $ 0 in the pure phases in the case of skyrmion host materials. In contrast, in the present case, within the pure helical magnetic phase a finite difference between the $ \chi_{S}$ and $\chi_{T} $ is found especially at lower magnetic fields. 

(iii) The presence of maxima in the $ \tau_{0}(H) $ at \textit{H} = $H_{1}$ and $H_{2}$ [see Fig. \ref{Fig3}(d)] indicates slower magnetization dynamics in the vicinity of the phase transformation that hints at a mixed phase of different magnetic structures. $ \tau_{0} $ at $H_{1}$ and $H_{2}$ is nearly 1.5 to 3 times higher in comparison to the corresponding values in the pure magnetic phases. In case of Fe$ _{1-x} $Co$ _{x} $Si \cite{L.J.Bannenberg},  Cu$_{2}$OSeO$_{3}$ \cite{I.Levatic,F.Qian} and GaV$_{4}$S$_{8}$ \cite{A.Butykai} the change in $ \tau_{0} $ is found to be as large as  in the order of 10$ ^{1} $ to 10$ ^{2} $. Slower relaxation phenomena characterized by higher relaxation time in the vicinity of phase boundaries signifies possible origin of irregularities in the spin-coordination between different magnetic textures.

(iv) The relaxation time distribution parameter, $ \alpha $, takes very small values in the range 0.05 - 0.15. Low values of $ \alpha $ suggest that the relaxation process of the configurations happens nearly in unison. In other words, topologically protected magnetic  antiskyrmions, whether in the long-range ordered lattice (Askx phase) phase or in the disordered state (H+Askx phase / FP+Askx phase), undergo almost simultaneous relaxation. Nevertheless, $ \alpha $ tends to rise at the phase crossover regions.

(v) Smaller values of  $ \tau_{0} \simeq$  10$ ^{-5} $ s, compared to that of 10$ ^{-3} $ s reported for the skyrmion host Cu$_{2} $OSeO$_{3}$, reflects faster relaxation in Mn$_{1.4} $PtSn, presumably due to the large spin-orbit coupling introduced by the presence of heavy element `Pt' in the material. In the absence of a quantitative proof, the premise that the strong spin-orbit coupling due to `Pt' is the cause for the faster relaxation times in the present system remains a conjecture only.

(vi) Temperature variation of the characteristic relaxation times corresponding to Helical to AskX and AskX to FP phase are plotted in the upper and lower panels of Fig. \ref{Fig3}(f). In general, a  field of 2 kOe covers the antiskyrmion phase in the whole range of temperatures  in the present study. Hence, the temperature dependence of the relaxation time at $ H $ = 2 kOe is shown in the central panel of the Fig. \ref{Fig3}(f). In all the cases, the relaxation behavior can be described using the well-known Arrhenius law $ \tau_{0} \propto  exp\left( E/k_{B}T\right) $, where $ E $ characterizes the energy barrier over which the thermal activation takes place. The fits based on the above relation are shown as continuous lines to the data presented in symbols. The fitted values yield activation energies (in the units of temperature, $ E/k_{B} $) of 626(30) K, 882(7) K, 226.5(7.0) K across the H-AskX, AskX, and AskX-FP phases, respectively. A higher value of the energy barrier in the core-antiskyrmion lattice region might have resulted from the strong topological protection. Lower values of \textit{E} at H-AskX, AskX-FP phase boundaries signals at metastable antiskyrmions, being protected by somewhat lesser energy barriers.

%%%%%%%%%%%%%%%%%%%%%%%%%%%%%%%%%%%%%%%%%%%%%%%%%%%%%%%%%%%%%%%%%%%%%%%%%%%%%%%%%%%%%%%%%%%%%%%%%%%%%%%%%%%%%%%%%%%%%%%%%%%%%%%%%%%%%%%%%%%%%%%%%%%%%
\begin{figure} [tb!]
	\includegraphics[angle=0,width=8.5cm,clip]{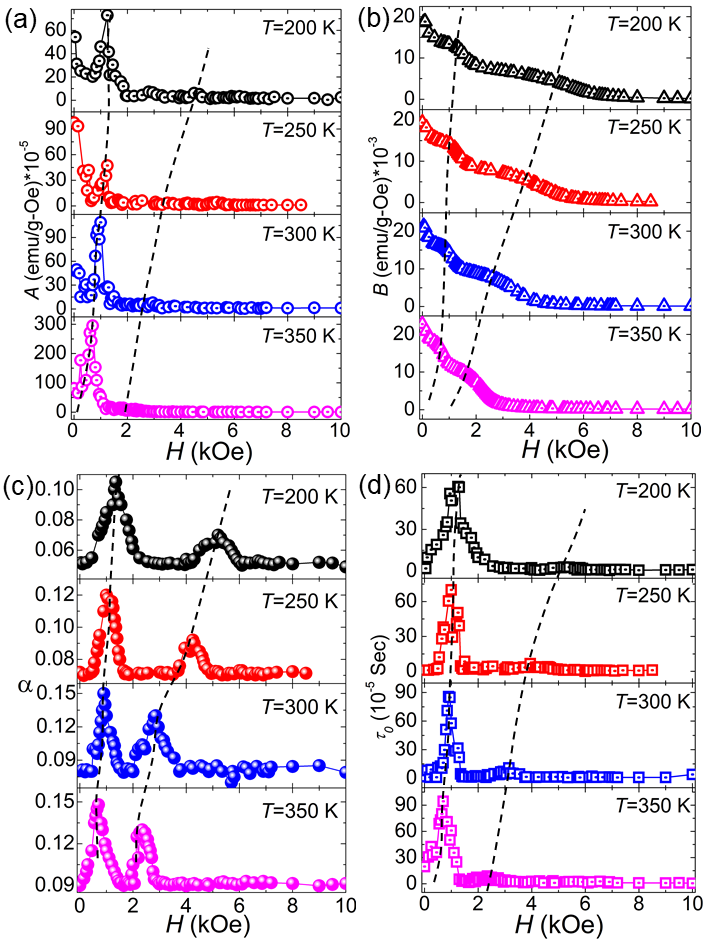}
	\caption{\label{FIG4}(Color online) 	Field dependent fitting parameters for Mn$_{1.4}$Pt$_{0.9}$Pd$_{0.1}$Sn (a) $ \chi_{T} - \chi_{S} $, (b) $B = \chi_{S} $,  and (e)  $ \alpha $ and (d) $ \tau_{0} $. Dashed lines serve as guide to eye. }
	\label{Fig4}
\end{figure}
%%%%%%%%%%%%%%%%%%%%%%%%%%%%%%%%%%%%%%%%%%%%%%%%%%%%%%%%%%%%%%%%%%%%%%%%%%%%%%%%%%%%%%%%%%%%%%%%%%%%%%%%%%%%%%%%%%%%%%%%%%%%%%%%%%%%%%%%%%%%%%%%%%

While the observation of antiskyrmion phase in Mn$_{1.4}$PtSn system was briefly mentioned earlier, a complete microscopic study was reported in case of Mn$_{1.4}$Pt$_{0.9}$Pd$_{0.1}$Sn  \cite{A.K.Nayak}. To further support the present analysis using the Cole-Cole formalism, we have carried out a similar measurement protocol in case of Mn$_{1.4}$Pt$_{0.9}$Pd$_{0.1}$Sn. The frequency dependence of $ \chi^{\prime}\left(f\right)$ and  $\chi^{\prime\prime}\left(f\right)$ data are fitted using the  Cole-Cole relation as discussed earlier \cite{Supp}. Various fit parameters obtained from the $ \chi^{\prime}\left(f\right)$ fitting are plotted in Fig. 4. As it can be seen, all the parameters $A$, $B$, $\alpha$, and  $ \tau_{0} $ display a well-defined peak at the low-field regime indicating the nucleation of antiskyrmion phase with application of magnetic fields. Similarly, a hump kind of behaviour is observed at higher fields for $A$, $B$ and  $ \tau_{0} $, and a clear peak nature is found in case of the relaxation time distribution parameter, $ \alpha $. In addition, all the fit parameters fall in the same range as that is realized in case of  Mn$_{1.4}$PtSn. All these findings further validate  the present method of characterizing antiskyrmion phase in the tetragonal Heusler materials.

%%%%%%%%%%%%%%%%%%%%%%%%%%%%%%%%%%%%%%%%%%%%%%%%%%%%%%%%%%%%%%%%%%%%%%%%%%%%%%%%%%%%%%%%%%%%%%%%%%%%%%%%%%%%%%%%%%%%%%%%%%%%%%%%%%%%%%%%%%%%%%%%%%%%%
\begin{figure} [tb!]
	\includegraphics[angle=0,width=8.5cm,clip]{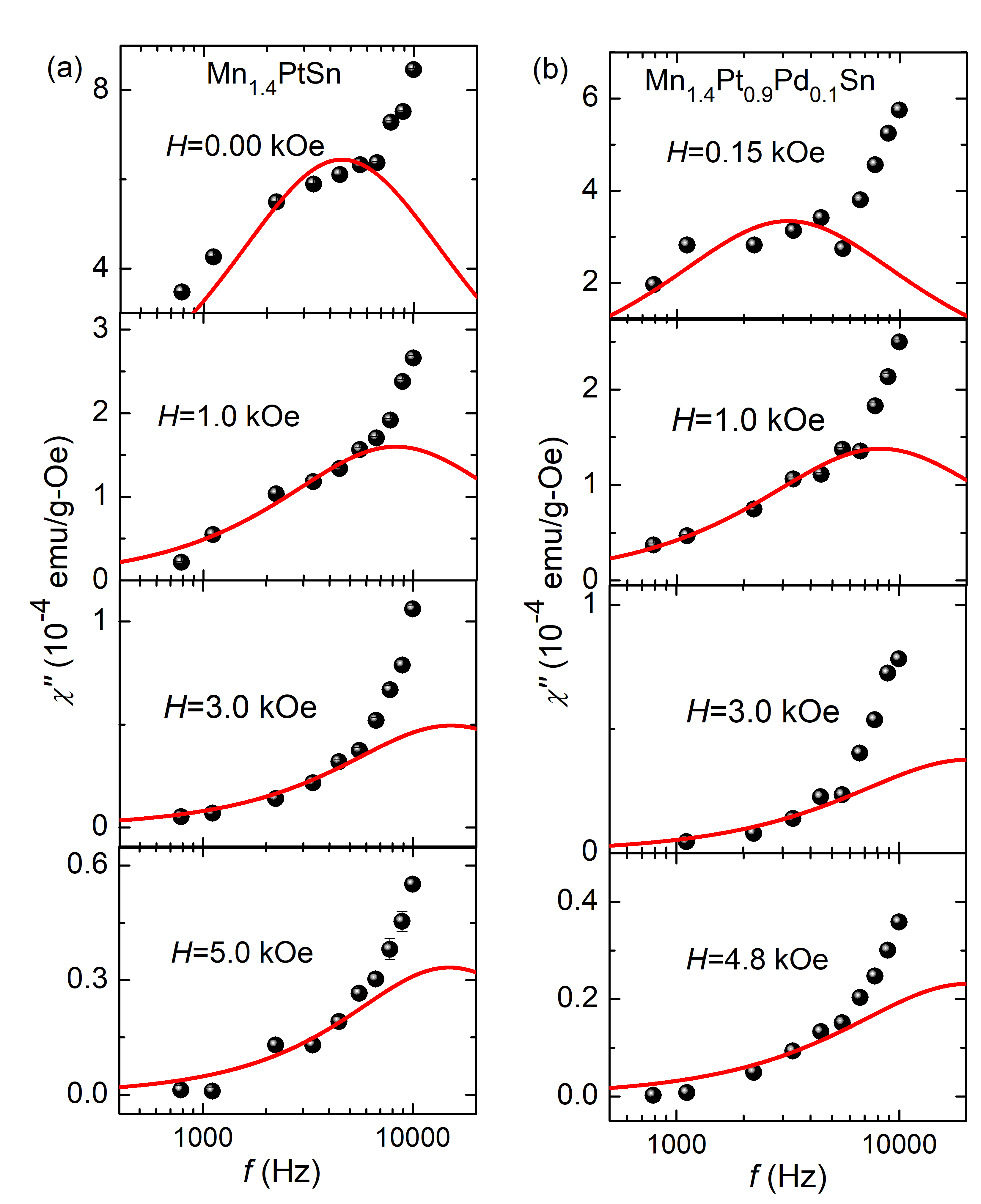}
	\caption{\label{FIG5}(Color online) 	Frequency dependence of $ \chi^{\prime\prime}$ at various representative magnetic fields measured at $ T $ = 300 K for (a) Mn$_{1.4}$PtSn and (b) Mn$_{1.4}$Pt$_{0.9}$Pd$_{0.1}$Sn. Solid balls represent experimental date points, whereas, the lines indicate fitting to the data using Eq. \ref{Eq.2}. }
	\label{Fig5}
\end{figure}
%%%%%%%%%%%%%%%%%%%%%%%%%%%%%%%%%%%%%%%%%%%%%%%%%%%%%%%%%%%%%%%%%%%%%%%%%%%%%%%%%%%%%%%%%%%%%%%%%%%%%%%%%%%%%%%%%%%%%%%%%%%%%%%%%%%%%%%%%%%%%%%%%%
 It is well-known that the out-of-phase component of the ac-susceptibility in the Debye model like relaxation shows a peak at a characteristic frequency ($ f_{0} = 1/\tau_{0} $). However, in the present case the $ \chi^{\prime\prime} (f)$ data show a shoulder nearly at $ f $ where $\chi^{\prime}(f)$ goes through an inflection point as shown in  Fig. \ref{Fig5}. A complete range of $ \chi^{\prime\prime} (f)$ data at different temperatures and magnetic fields can be found in the supplementary information \cite{Supp}.  We have attempted to fit the $ \chi^{\prime\prime}\left(f\right)$ data using Eq. \ref{Eq.2} as depicted by solid lines in Fig. \ref{Fig5}.   We use the fit parameters obtained from the $\chi^{\prime}(f)$ analysis as inputs for the fitting of $ \chi^{\prime\prime} (f)$ data. Then the parameters were relaxed to obtain the best fitting.  However, we could not fit the data in the whole frequency range owing to the lack of complete peak kind of behaviour, in the case of both Mn$_{1.4}$PtSn and  Mn$_{1.4}$Pt$_{0.9}$Pd$_{0.1}$Sn. A good match between the experimental data and the fitted curve is obtained only up to the frequency where $ \chi^{\prime\prime} (f)$ shows a shoulder like behavior. This is true for all the temperatures and fields covered in the present study \cite{Supp}. The fit parameters obtained from the $ \chi^{\prime\prime} (f)$ analysis fall in the same order of magnitude as that found from the $ \chi^{\prime} (f)$ fitting. A comparison of fit parameters acquired from the $ \chi^{\prime} (f)$ and  $ \chi^{\prime\prime} (f)$ fittings at 300 K for both Mn$_{1.4}$PtSn and  Mn$_{1.4}$Pt$_{0.9}$Pd$_{0.1}$Sn is presented in Table I. Interestingly, both the samples exhibit similar magnitude of the respective fit parameters, suggesting presence of similar relaxation dynamics in the present systems. This can be understand from the fact that both the samples display identical magnetic ordering with almost equal $ T_C $.
 
%%%%%%%%%%%%%%%%%%%%%%%%%%%%%%%%%%%%%%%%%%%%%%%%%%%%%%%%%%%%%%%%%%%%%%%%%%%%%%%%%%%%%%%%%%%%%%%%%%%%%%%%%%%
\begin{table}
	\caption {\label{tab:table1} Comparison between fitted parameters of $\chi^{\prime}$ and $ \chi^{\prime\prime}$ at 300 K.}

	\begin{tabular}{l||l l|l l|l l}
		\hline \hline
		\multirow{2}{*}{H(Oe)}   & \multicolumn{2}{l||}{A($\times$ 10$^{-5}$ emu/g-Oe)} & \multicolumn{2}{l||}{$\tau_0$( $\times$ 10$^{-5}$ s)} & \multicolumn{2}{l}{$\alpha$} \\ \cline{2-7} 
		& $\chi^{\prime}$ & $ \chi^{\prime\prime}$        & $\chi^{\prime}$         & $ \chi^{\prime\prime}$    & $\chi^{\prime}$          & $ \chi^{\prime\prime}$         \\ \hline\hline
		\multirow{2}{*}{0}\hspace{0.57cm}$^{\mathrm{a}}$   & 12.220                     $\space$ $\space$ $\space$$\space$$\space$& 149.000         & 5.070       $\space$ $\space$ & 3.500         & 0.042          $\space$ $\space$& 0.092         \\ \cline{2-7} 
		\hspace{0.70cm}$^{\mathrm{b}}$ & 54.193          & 122.358        & 1.536        & 6.623        &   0.081         &     0.080      \\ \hline
		\multirow{2}{*}{500}\hspace{0.20cm} $^{\mathrm{a}}$ & 7.379         &  27.071         &   6.943        &   3.060         &    0.057        &     0.045      \\ \cline{2-7} 
		\hspace{0.65cm} $\space$$^{\mathrm{b}}$ &     15.122       &  40.136         &  10.979         &   3.501         &   0.103         &  0.083         \\ \hline
		\multirow{2}{*}{1000}$\space$ $^{\mathrm{a}}$    & 8.051           &   36.280        &    8.287       &     1.930       &  0.048          &   0.080        \\ \cline{2-7} 
		\hspace{0.65cm} $^{\mathrm{b}}$& 87.357          &  26.957         &   57.727        &   2.359         &  0.130          & 0.082          \\ \hline
		\multirow{2}{*}{1500}$\space$ $^{\mathrm{a}}$    &  7.340         & 25.208           &  8.671         &  1.682          &  0.010          & 0.128          \\ \cline{2-7} 
		\hspace{0.65cm} $^{\mathrm{b}}$& 10.476          & 12.345          & 1.480          &  1.214          &   0.086         &    0.081       \\ \hline
		\multirow{2}{*}{2000}$\space$ $^{\mathrm{a}}$    & 3.784           &  17.365         &  5.198         &    1.185        &  0.036          &    0.034       \\ \cline{2-7} 
		\hspace{0.65cm} $^{\mathrm{b}}$&   5.337         &    10.251       &  1.944         &  0.946          &  0.107          &   0.080        \\ \hline
		\multirow{2}{*}{2500}$\space$ $^{\mathrm{a}}$   & 1.873           & 13.680          &    3.880       &   1.006         & 0.020           &  0.051         \\ \cline{2-7} 
		\hspace{0.65cm} $^{\mathrm{b}}$&  3.998          &  8.724         &   2.516       &    0.872       &   0.104         &     0.083      \\ \hline
		\multirow{2}{*}{3000}$\space$ $^{\mathrm{a}}$   & 2.761           &  10.721         & 5.437          &  1.048          &  0.040          &     0.051      \\ \cline{2-7} 
		\hspace{0.65cm} $^{\mathrm{b}}$&    7.762        &   8.969        &   6.971        &     0.989       &   0.121         &    0.082       \\ \hline
		\multirow{2}{*}{3500}$\space$ $^{\mathrm{a}}$   & 2.863           & 9.767           &  4.567         &  1.301         &   0.090         &     0.040      \\ \cline{2-7} 
		\hspace{0.65cm} $^{\mathrm{b}}$& 3.803           &   8.113        &   4.383        &   0.824         &  0.092          & 0.088          \\ \hline
		\multirow{2}{*}{4000}$\space$ $^{\mathrm{a}}$   &  2.835          &  7.383         &  3.389        &   1.222        &   0.020         &  0.070         \\ \cline{2-7} 
		\hspace{0.65cm} $^{\mathrm{b}}$& 2.224         &    6.138       &    1.314       &  0.805          &  0.084          &   0.085        \\ \hline
		\multirow{2}{*}{5000}$\space$ $^{\mathrm{a}}$   &  2.584          &  6.873         &   0.628        &  1.067          & 0.010           &   0.020        \\ \cline{2-7} 
		\hspace{0.65cm} $^{\mathrm{b}}$& 1.253           &  6.297         &   1.470        &  0.854          &  0.082          &  0.084         \\ \hline\hline
	\end{tabular}
	\begin{tabbing}
		$^{\mathrm{a}}$Mn$_{1.4}$PtSn, 
		$^{\mathrm{b}}$Mn$_{1.4}$Pt$_{0.9}$Pd$_{0.1}$Sn
	\end{tabbing}
\end{table}

%%%%%%%%%%%%%%%%%%%%%%%%%%%%%%%%%%%%%%%%%%%%%%%%%%%%%%%%%%%%%%%%%%%%%%%%%%%%%%%%%%%%%%%%%%%%%%%%%%%%%%%%%%%

 The deviation of the present $ \chi^{\prime\prime} (f)$ data from the Debye model like relaxation could stem from the eddy current losses that adversely affect the susceptibility signals, particularly at higher frequencies and higher dc fields. In general, the voltage induced in the pick-up coil by the eddy currents generated in the sample is much smaller when compared to the main voltage induced from the sample in response to the applied ac magnetic field. As a result, the eddy current effect can be clearly seen when the actual signal is very weak \cite{Kraftmakher}.  In the present case, the eddy current effect is predominant in case of $ \chi^{\prime\prime} (f)$  as the signal of magnitude $\sim  10^{-5} - 10^{-6}$ emu/Oe falls in the sensitivity limit of the measuring instrumen. Similar magnitude of the eddy current effect of the order $\sim  10^{-7}$$  m^3 $/mol or $\sim  10^{-5} $ emu/Oe is also found in the $ \chi^{\prime\prime} (f)$ data at 1 kHz frequency in case of the semiconducting skyrmion host Fe$ _{1-x} $Co$_ x $Si \cite{L.J.Bannenberg}. The eddy current effect in the previously studied skyrmion hosting materials could not be seen prominantly due to the fact that the magnetic relaxation in these systems happens in low frequency regime spanning up to 1 kHz, whereas, the present antiskyrmion materials require frequencies up to 10 kHz to study the ralaxation process.

 %%%%%%%%%%%%%%%%%%%%%%%%%%%%%%%%%%%%%%%%%%%%%%%%%%%%%%%%%%%%%%%%%%%%%%%%%%%%%%%%%%%%%%%%%%%%
 \begin{table}[b]
 	\centering
 	\caption{\label{tab:table2}Comparison of relaxation times  $\tau_{0}$  in different magnetic systems.}
 	\begin{ruledtabular}
 		\begin{tabular}{cccc}
 			\begin{tabular}{@{}c@{}}Magnetic system \end{tabular} &  $\tau_{0}$ (s) \\
 			\hline
 			Canonical/Cluster spin glasses{\cite{J.A.Mydosh}} & 10$^{-12}$ - 10$^{-8}$ \\ 
 			Superparamagnets{\cite{S.J.Blundell2}} & 10$^{-11}$ - 10$^{-9}$ \\ 
 			Single molecular magents{\cite{L.Thomas}} & 10$^{-6}$ \\ 
 			Skyrmion-Cu$_{2}$OSeO$_{3}${\cite{F.Qian}}(50 $ \pm  $ 10 nm) & 10$^{-3}$ - 10$^{-2}$ \\
 			Antiskyrmion-Mn$_{1.4}$Pt(Pd)Sn (150 $ \pm  $ 20 nm) & 10$^{-5}$ - 10$^{-4}$ \\ 
 		\end{tabular}
 	\end{ruledtabular}
 \end{table}
%%%%%%%%%%%%%%%%%%%%%%%%%%%%%%%%%%%%%%%%%%%%%%%%%%%%%%%%%%%%%%%%%%%%%%%%%%%%%%%%%%%%%%%%%%%%%%%%%%%%%%%%%%%
It is customary here to compare various magnetic systems that display non-trivial and finite-sized magnetic configurations such as spin glasses, superparamagnetic particles, single molecular magnets, and skyrmion lattices that exhibit slow magnetic relaxations and characterized by Debye relation. Table \ref{tab:table2} compares the typical relaxation times in such systems. In case of spin glasses, the reorientation of individual magnetic moments or clusters during the relaxation process causes additional frustrations. This   result in further rearrangements of the local spins in the system leading to slower magnetization dynamics. In superparamagnets and single molecular magnets, slow magnetic relaxation arises from the thermally-activated hopping of the spins(macro) over energy barriers dictated by the uniaxial anisotropy which separates parallel and anti-parallel configurations. A phase crossover from helical spin modulations to skyrmion/antiskyrmion lattice associates with topological point defects \cite{X.Z.Yu}. It is reported that the reorientation of large magnetic entities, such as modulations in the long-wavelength spin helices, nucleation of skyrmion cores, and glassy nature of the skyrmion phases \cite{Rajeswari} leads to slower magnetization dynamics in different skyrmion hosts \cite{A.Bauer,L.J.Bannenberg,I.Levatic,F.Qian,A.Butykai}. Similar phenomena are also expected in the present antiskyrmion host Mn-Pt(Pd)-Sn materials. Higher relaxation time ($ \tau_{0}  \approx  10^{-3} - 10^{-5} $ s), i.e., slower relaxation process in the skyrmion/antiskyrmion systems when compared to others signifies slower magnetic damping process,  which suppresses the rate at which an equilibrium configuration can be restored. In this regard, resonance studies of the magnetization dynamics can help to understand the stability and reliability of future skyrmion/antiskyrmion-based spintronics.

%%%%%%%%%%%%%%%%%%%%%%%%%%%%%%%%%%%%%%%%%%%%%%%%%%%%%%%%%%%%%%%%%%%%%%%%%%%%%%%%%%%%%%%%%%%%%%%%%%%%%%%%%%%%%%%%%%%%%%%%%%%%%%%%%%%%%%%%%%%%%%%%%%%%%
\begin{figure} [tb!]
	\includegraphics[angle=0,width=8.5cm,clip]{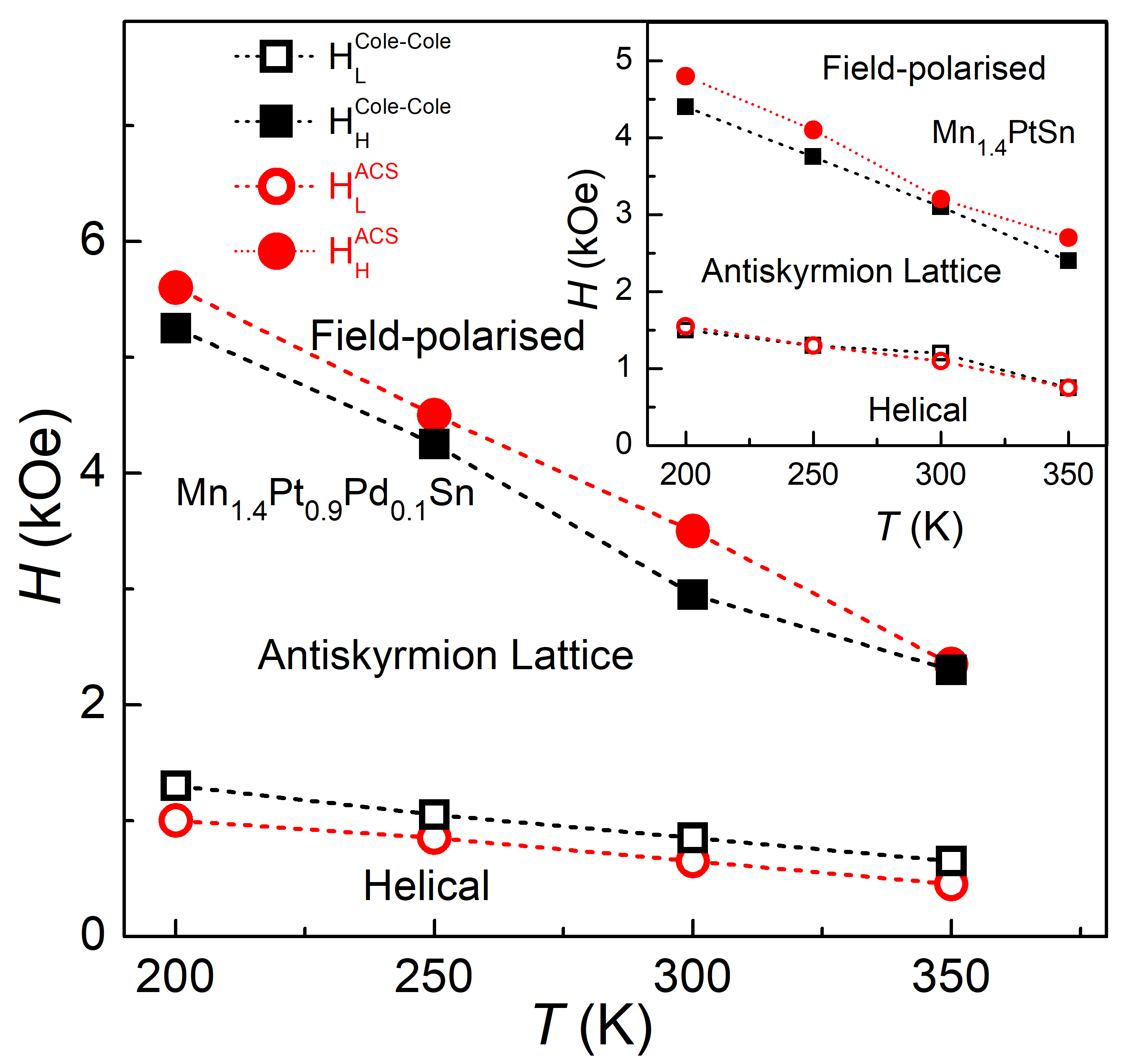}
	\caption{\label{FIG6}(Color online) $ H-T$ phase diagram inferred from maxima in the ac susceptibility (circles) and  the fit parameters of the Cole-Cole formalism (Squares) as described in the text for Mn$_{1.4}$Pt$_{0.9}$Pd$_{0.1}$Sn. The inset shows $ H-T$ phase diagram for Mn$_{1.4}$PtSn.  Open and closed  symbols represent the lower and upper boundaries of antiskyrmion phase, respectively.  $ H_{L}^{ACS} $ and $ H_{H}^{ACS} $ show the lower and upper phase boundaries of the antiskyrmion phase, respectively, obtained from the peak/hump anomalies in the field-evaluation of the ac susceptibility measurements. Similarly, $ H_{L}^{Cole-Cole} $ and $ H_{H}^{Cole-Cole} $ are lower and upper phase boundaries of the antiskyrmion phase obtained from the field-dependent fit parameters of the Cole-Cole expression using $ \chi^{\prime}(H) $ data.}
	\label{Fig6}
\end{figure}
%%%%%%%%%%%%%%%%%%%%%%%%%%%%%%%%%%%%%%%%%%%%%%%%%%%%%%%%%%%%%%%%%%%%%%%%%%%%%%%%%%%%%%%%%%%%%%%%%%%%%%%%%%%%%%%%%%%%%%%%%%%%%%%%%%%%%%%%%%%%%%%%%%

%%%%%%%%%%%%%%%%%%%%%%%%%%%%%%%%%%%%%%%%%%%%%%%%%%%%%%%%%%%%%%%%%%%%%%%%%%%%%%%%%%%%%%%%%%%%%%%%%%%%%%%%%%%%%%%%%%%%%%%%%%%%%%%%%%%%%%%%%%%%%%%%%%%%%%

$H$-$T$ phase diagrams are established from the above-discussed features of $ \chi^{\prime}(H) $ and the field-dependent fit parameters of the Cole-Cole relation on the polycrystalline samples Mn$_{1.4}$Pt$_{0.9}$Pd$_{0.1}$Sn and Mn$_{1.4}$PtSn as shown in Fig. \ref{Fig6}. It may be noted here that the $ H_{L}^{ACS} $($ H_{L}^{Cole-Cole} $) and $ H_{H}^{ACS} $($ H_{H}^{Cole-Cole} $) are the lower and upper phase boundaries of the antiskyrmion phase which describe the crossover from the helical state to the antiskyrmion region and the antiskyrmion region to the field-polarized state, respectively. The phase boundaries obtained from the Cole-Cole analysis  in the present system broadly matches with that estimated from the magnetic entropy study \cite{S.K.Jamal} and the Lorentz transmission electron microscopy imaging (LTEM) technique \cite{A.K.Nayak}. The discrepancy seen in the variation of the upper critical fields at low temperature regime mostly originates from the non-identical orientations of the crystallites in the polycrystalline samples used for different measurements. This is due to the fact that the angle between the applied magnetic field and the crystallographic orientation of the crystallites determines the critical field required for the stabilization of antiskyrmions \cite{A.K.Nayak}. Hence, depending upon the orientation of the  crystallites the critical field for the nucleation of antiskyrmions may vary for different pieces cut from the same sample. The discrepancy found for the   lower critical fields in the present study to that of LTEM investigation \cite{A.K.Nayak} can be understood as follows. In the phase diagram inferred from the LTEM studies, typically magnetic field was applied at room temperature to stabilize the antiskyrmion lattice and then \textit{`field-cooled'} the sample to the required temperatures. Subsequently, the magnetic field was slowly varied in different steps, i.e., either reduced to zero or increased to obtain a field-polarized state and simultaneously monitored the emergence of different magnetic phases. Whereas, in the present case, the sample was \textit{`zero-field-cooled'} from 400~K to the required temperature and the magnetic field strength is continuously raised starting from a zero value to see the field evolution of helical, antiskyrmion and field-polarized phases. In addition, the Mn-Pt(Pd)-Sn samples used in the LTEM study \cite{A.K.Nayak} are [001] oriented 100 nm thin-crystalline films prepared by the Focused Ion Beam (FIB) technique. Whereas, the present ac-susceptibility measurements are carried out on the polycrystalline bulk samples. Therefore, it is expected that the phase diagram for the bulk polycrystalline samples may differ slightly in comparison to the single grain thin-crystalline-film sample. 

%%%%%%%%%%%%%%%%%%%%%%%%%%%%%%%%%%%%%%%%%%%%%%%%%%%%%%%%%%%%%%%%%%%%%%%%%%%%%%%%%%%%%%%%%%%%%%%%%%%%%%%%%%%%%%%%%%%%%%%%%%%%%%%%%%%%%%%%%%%%%%%%%%%%%

\section{CONCLUSION}
In summary, we have presented  a detailed study on the frequency-dependent ac-susceptibility measurements on the  $D_{2d}$ symmetry based tetragonal  Heusler system that hosts antiskyrmion lattice over a wide temperature range. The magnetic relaxation follows a Debye relation with nearly negligible distribution in the relaxation times. At the boundaries between the different phases \textit{via} helical, antiskyrmion, and field-polarized states, the characteristic relaxation time, the isothermal and adiabatic susceptibilities show a non-monotonic variation. Maxima in the relaxation times that develop in the vicinity of the phase crossover indicate slower magnetization dynamics. Temperature-dependent relaxation times across different magnetic phases follow an Arrhenius kind of slowing down with reasonable values of the energy barriers. The observation of  higher relaxation times in comparison to other particulate storage mediums such as super-paramagnets, and single molecular magnets  signifies slower  damping process in the present antiskyrmion host Mn-Pt(Pd)-Sn system.

% %%%%%%%%%%%%%%%%%%%%%%%%%%%%%%%%%%%%%%%%%%%%%%%%%%%%%%%%%%%%%%%%%%%%%%

\begin{acknowledgments}
		This work was financially supported by the Science and Engineering Research Board (SERB) under research grant (No. ECR/2017/000845), Department of Science and Technology (DST)-Ramanujan research grant (No. SB/S2/RJN-081/2016) and  Nanomission research grant  [SR/NM/NS-1036/2017(G)] of the Government of India. AKN acknowledges the Max Plank Society for support under the Max Plank-India partner group project.
	
P.V.P.M. and S.S. contributed equally to this work.	
	
\end{acknowledgments}

% %%%%%%%%%%%%%%%%%%%%%%%%%%%%%%%%%%%%%%%%%%%%%%%%%%%%%%%%%%%%%%%%%%%%%%

%%%%%%%%%%%%%%%%%%%%%%%%%%%%%%%%%%%%%%%%%%%%%%%%%%%%%%%%%%%%%%%%%%%%%%%%%%%%%%%%%%%%%%%%%%%%%%%%%%%%%%%%%%%%%%%%%%%%%%%%%%%%%%%%%%%%%%%%%%%%%%%%%%
\section{SUPPLEMENTARY INFORMATION}
\subsection{Experimental methods: Sample preparation and Structural Characterization}

Polycrystalline ingots of Mn$_{1.4}$PtSn, Mn$_{1.4}$Pt$_{0.9}$Pd$_{0.1}$Sn were prepared by arc melting technique. The constituent elements were taken according to their stoichiometric ratio and melted together into an ingot (under high pure argon atmosphere) inside the Arc-melting furnace. For better homogeneity, the ingots were melted several times by flipping them upside down. Because of volatile nature of manganese, some extra amount of Mn was added to compensate the weight loss during melting. As the ingots are practically quenched after the melting, annealing of arc melted ingots is mandatory for better homogeneity and phase formation. Hence the arc-melted ingots were sealed inside quartz ampoule under ~10$ ^{-5} $ mbar pressure. The quartz tube containing the sample was kept in PID-controlled resistive furnace for heat treatment at 1073 K for 7 days. Subsequently, the samples were quenched in ice-water mixture. Structural characterization and phase purity conformation of the annealed sample were characterized by the x-ray powder diffraction (XRD) using a Cu-$ K_{\alpha} $ source. \\

Room-temperature x-ray diffraction pattern along with the Rietveld refinement fit is shown in Fig.7. It is observed that Mn$_{1.4}$PtSn crystallizes to a non-centrosymmetric tetragonal lattice with space group I$\bar{4}$ 2m (space group no: 121). The lattice parameters obtained from the Rietveld analysis are $a$ = $b$ = 6.3478 $ \pm $ 0.0005 Å and $c$ = 12.2006 $  \pm$ 0.0007 Å with $ c/a $ = 1.9220. The sample is polycrystalline in nature with crystallite sizes of the order of several tens of microns. For this reason, when we tried to find out the average crystallite size from the Rietveld refinement using the instrument resolution file (IRF), we obtained negligible size broadening.\\

%%%%%%%%%%%%%%%%%%%%%%%%%%%%%%%%%%%%%%%%%%%%%%%%%%%%%%%%%%%%%%%%%%%%%%%%%%%%%%%%%%%%%%%%%%%%%%%%%%%%%%%%%%%%%%%%%%%%%%%%%%%%%%%%%%%%%%%%%%%%%%%%%%%%%
\begin{figure}
	\includegraphics[angle=0,width=8.5cm,clip]{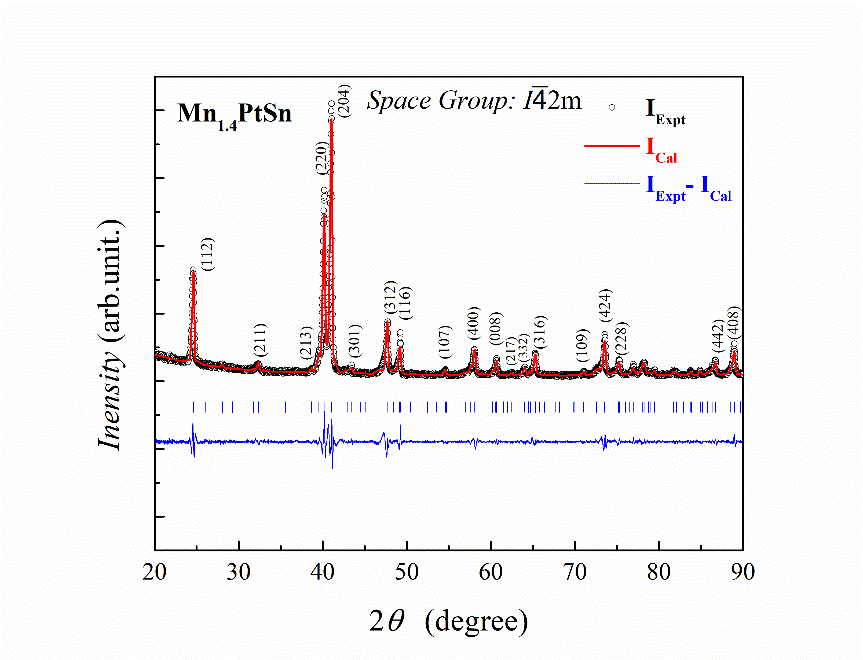}
	\caption{\label{FIG7}(Color online) Room temperature powder XRD pattern for  Mn$_{1.4}$PtSn.}
	\label{Fig7}
\end{figure}
%%%%%%%%%%%%%%%%%%%%%%%%%%%%%%%%%%%%%%%%%%%%%%%%%%%%%%%%%%%%%%%%%%%%%%%%%%%%%%%%%%%%%%%%%%%%%%%%%%%%%%%%%%%%%%%%%%%%%%%%%%%%%%%%%%%%%%%%%%%%%%%%%%
Compositional homogeneity of the MnPtSn sample was confirmed by Energy-dispersive x-ray spectroscopy (EDX) analysis (See figure 8). The EDX analysis yields a composition of 41.14 $ \pm $ 1.50 $ at. \% $ for Mn, 28.36 $ \pm $ 1.00 $ at. \% $ for Pt and 30.50 $ \pm $ 1.00 $ at. \% $ for Sn, finally resulting a composition of  Mn$_{1.4}$PtSn.
%%%%%%%%%%%%%%%%%%%%%%%%%%%%%%%%%%%%%%%%%%%%%%%%%%%%%%%%%%%%%%%%%%%%%%%%%%%%%%%%%%%%%%%%%%%%%%%%%%%%%%%%%%%%%%%%%%%%%%%%%%%%%%%%%%%%%%%%%%%%%%%%%%%%%
\begin{figure}
	\includegraphics[angle=0,width=8.5cm,clip]{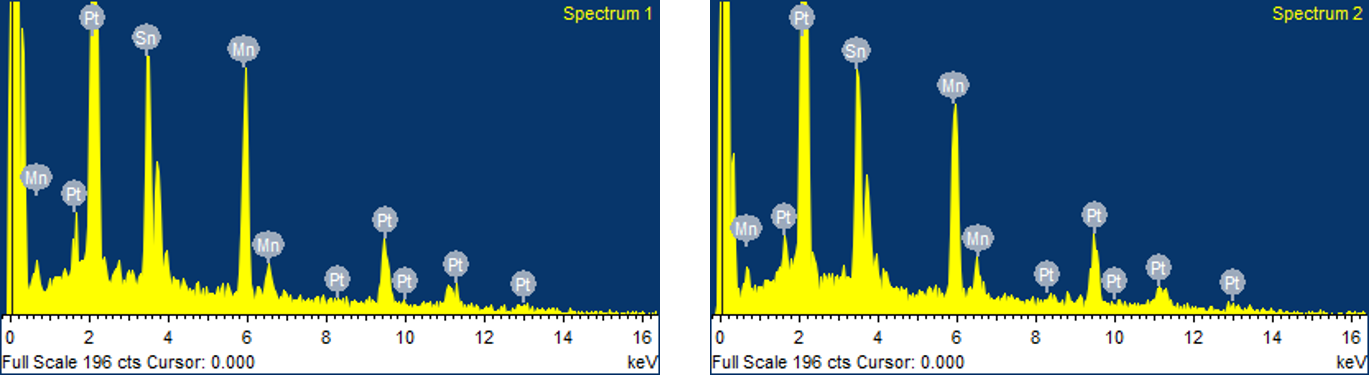}
	\caption{\label{FIG8}(Color online) EDX spectrums measured on Mn$_{1.4}$PtSn sample.}
	\label{Fig8}
\end{figure}
%%%%%%%%%%%%%%%%%%%%%%%%%%%%%%%%%%%%%%%%%%%%%%%%%%%%%%%%%%%%%%%%%%%%%%%%%%%%%%%%%%%%%%%%%%%%%%%%%%%%%%%%%%%%%%%%%%%%%%%%%%%%%%%%%%%%%%%%%%%%%%%%%%
\subsection{AC susceptibility at frequencies in the range 11 Hz $ \leq $ f $ \leq $ 666 Hz}
Fig.9 testifies that the $ \chi^{\prime} $(H, T = 300 K) and $ \chi^{\prime\prime} $(H, T = 300 K) at frequencies ($ \leq $ 500 Hz) are almost independent of frequency of the \textit{ac} signal. This is also true at other temperatures. Note that, the $ \chi^{\prime\prime} $ signal is nearly an order of magnitude smaller when compared to that of at higher frequencies and falls in the instrument resolution limit, leading to the scattering in the data.  
%%%%%%%%%%%%%%%%%%%%%%%%%%%%%%%%%%%%%%%%%%%%%%%%%%%%%%%%%%%%%%%%%%%%%%%%%%%%%%%%%%%%%%%%%%%%%%%%%%%%%%%%%%%%%%%%%%%%%%%%%%%%%%%%%%%%%%%%%%%%%%%%%%%%%
\begin{figure}
	\includegraphics[angle=0,width=8.5cm,clip]{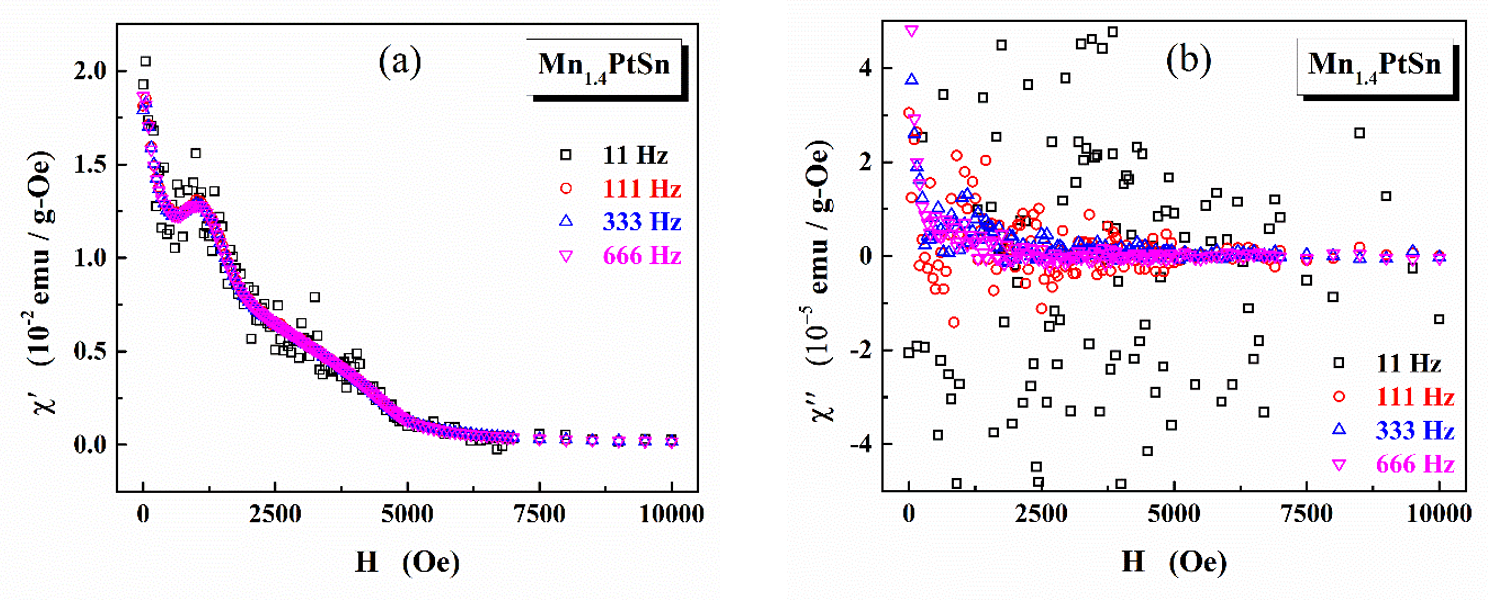}
	\caption{\label{FIG9}(Color online) Field-dependent (a) Real part, $ \chi^{\prime} $ and (b) imaginary part, $ \chi^{\prime\prime} $ of the susceptibilities at various ac-field frequencies in the range 11 Hz $ - $ 666 Hz measured at 300 K.}
	\label{Fig9}
\end{figure}
%%%%%%%%%%%%%%%%%%%%%%%%%%%%%%%%%%%%%%%%%%%%%%%%%%%%%%%%%%%%%%%%%%%%%%%%%%%%%%%%%%%%%%%%%%%%%%%%%%%%%%%%%%%%%%%%%%%%%%%%%%%%%%%%%%%%%%%%%%%%%%%%%%
\subsection{Theoretical considerations}
When probing a system with an oscillating magnetic field of certain frequency which is comparable to the characteristic frequency (equivalently timescale) of the magnetic relaxation of the system, then there will be some phase lag and hence dissipation. In such a case the total response ($ \chi $) can be conveniently expressed by the sum of in-phase, $ \chi^{\prime} $(real) and out-of-phase, $ \chi^{\prime\prime} $ (imaginary) components as 
\begin{equation}
 \chi =  \chi^{\prime}+i\chi^{\prime\prime}
 \label{Eq.3}
\end{equation}
Where the $ \chi^{\prime} $ reflects the sensitivity of a material to the applied field, also known as reversible magnetization, $ \chi^{\prime\prime} $ represents the dissipation of absorbed energy from the high frequency ac field i.e., irreversible magnetization \cite{A.H.Morrish}. This irreversibility results from the relaxation processes of various origin such as irreversible movement of domain walls, hysteresis loss in ferromagnets, magnetic phase transition, spin reorientation, spin-spin relaxation, spin-lattice relaxation to name a few. Since relaxation is very much sensitive to the magnetic phases, study of relaxation provides an insight into the magnetic structures, spin-spin and spin-lattice interactions in a material \cite{A.H.Morrish}. By considering analogy to the well-known Debye model of dielectric relaxation \cite{P.Debye}, Casimir and du Pr\'{e} \cite{Casimir} derived a thermodynamic model of relaxation for a magnetic system according to which the expression for complex susceptibility is given by,
\begin{equation}
 \chi(\omega) =  \chi_{S}+ \frac{\chi_{T}-\chi_{S}}{1+(i \omega \tau_{0})}
 \label{Eq.4}
\end{equation}
Where $ \chi_{T} \equiv \chi(\omega\longrightarrow 0) $ represents isothermal susceptibility in the limit of the lowest
frequencies, $ \chi_{S} \equiv \chi(\omega\longrightarrow \infty) $ is the adiabatic susceptibility in the limit of the highest frequency for which the spin system remains isolated from surroundings and $ \tau_{0} $ is the characteristic relaxation time constant and $ \omega = 2 \pi f $. The above Eq. (4) is called the ‘Debye relation’. The real ($ \chi^{\prime} $) and imaginary ($ \chi^{\prime\prime} $) components of the ac susceptibility from Eq. (4) can be written as,
\begin{subequations}
\begin{align}
 \chi^{\prime}(\omega) & =  \chi_{S}+ \frac{\chi_{T}-\chi_{S}}{1+(\omega^{2}\tau_{0}^{2})} \\ 
 \chi^{\prime\prime}(\omega) & =  (\chi_{T}-\chi_{S})+ \frac{\omega \tau}{1+(\omega^{2}\tau_{0}^{2})}
\end{align}
 \label{Eq.5}
\end{subequations}
A frequency dependent $ \chi^{\prime} $ and a non-zero value of $ \chi^{\prime\prime} $ are always consequence of one or more relaxation processes with a characteristic relaxation time constant ($ \tau_{0} $). Eq. (4) represents the exponential time dependence of the relaxation with the assumption that the magnetization is a slowly relaxing ‘entity’ with a single time scale. But when the entities interact with each other it can produce clustering effects which lead to a distribution in the relaxation times. When a single relaxation no longer governs the system dynamics, interactions in the spin systems and presence of cooperative effects leads to a spread in the relaxation times. This is often accounted for by introducing a phenomenological parameter ‘$  \alpha$’ in the Eqn. (4) as,
\begin{equation}
 \chi(\omega) =  \chi_{S}+ \frac{\chi_{T}-\chi_{S}}{1+(i \omega \tau_{0})^{1-\alpha}}
 \label{Eq.6}
\end{equation}
The resulting expression represents a generalized Debye model, also known as Cole-Cole relation \cite{Cole.Cole}. A zero value of ‘$  \alpha$’ signifies a single relaxation process, while close to unity represents an infinitely broad distribution of relaxation times. The frequency dependence of the real and imaginary components from Eq.(6) can be explicitly written as

\begin{subequations}
\begin{align}
\chi^{\prime}(f) & = \chi_{S}+\frac{A~\left[1+\left(2\pi f\tau_{0}\right)^{1-\alpha} \sin\left( \dfrac{\pi\alpha}{2}\right) \right] }{1+2\left(2\pi f\tau_{0}\right)^{1-\alpha} \sin \left( \dfrac{\pi\alpha}{2}\right)+\left(2\pi f\tau_{0}\right)^{2\left( {1-\alpha} \right)} } \\
\chi^{\prime\prime}(f)& = \frac{A~\left[\left(2\pi f\tau_{0}\right)^{1-\alpha} \cos\left( \dfrac{\pi\alpha}{2}\right) \right] }{1+2\left(2\pi f\tau_{0}\right)^{1-\alpha} \sin \left( \dfrac{\pi\alpha}{2}\right)+\left(2\pi f\tau_{0}\right)^{2\left( {1-\alpha} \right)} }\label{Eq.7}
\end{align}
\end{subequations}

Where A = ($ \chi_{T} $ - $ \chi_{S} $ ) and $ \tau_{0} $ is the average relaxation time. Fits have been attempted to the $ \chi^{\prime}(f) $ data at different but fixed magnetic fields over the temperature range 200 K-350 K, based on the eq.(7a) keeping the A, $ \chi_{S} $, $ \tau_{0} $, $ \alpha $ as free fitting parameters. The fits (continuous lines) best
represent the measured data (symbols) as shown in the figures 10-13 for Mn$ _{1.4} $PtSn and 14-17 for Mn$ _{1.4} $Pt$ _{0.9} $Pd$ _{0.1} $Sn. Similarly, imaginary part of ac susceptibility data [$ \chi^{\prime\prime}(f) $] as a function frequency is fitted using the eq.(7b) by taking the same values of parameter as initial values that are used to fit $ \chi^{\prime}(f) $ data (as discussed in the main texts) as shown in figures 18-21 for Mn$ _{1.4} $PtSn and 22-25 for Mn$ _{1.4} $Pt$ _{0.9} $Pd$ _{0.1} $Sn.

%%%%%%%%%%%%%%%%%%%%%%%%%%%%%%%%%%%%%%%%%%%%%%%%%%%%%%%%%%%%%%%%%%%%%%%%%%%%%%%%%%%%%%%%%%%%%%%%%%%%%%%%%%%%%%%%%%%%%%%%%%%%%%%%%%%%%%%%%%%%%%%%%%%%%
\begin{figure*}[b]
	\includegraphics[width=14cm,clip]{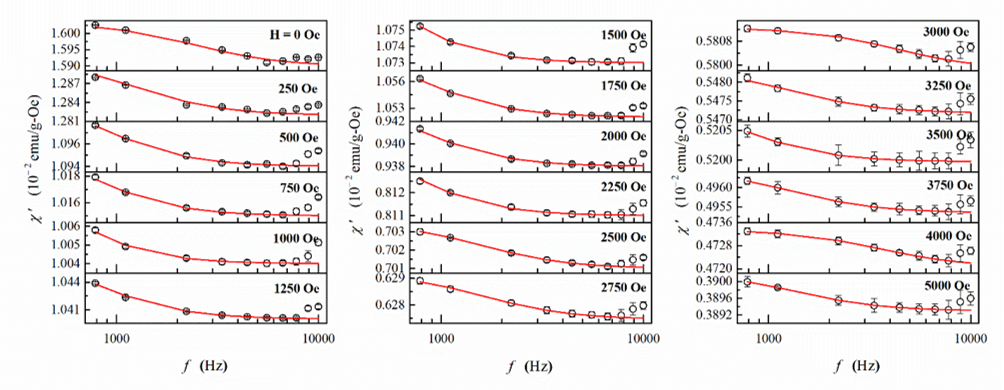}
	\caption{\label{FIG10}(Color online) Frequency dependence of $ \chi^{\prime} $ at various representative magnetic fields (H) in the case of T = 200 K for Mn$ _{1.4} $PtSn. The fits (solid lines) to the data (open symbols) are based on the Eq.(7a) described in the text.}
	\label{Fig10}
\end{figure*}
\begin{figure*}
	\includegraphics[width=14cm,clip]{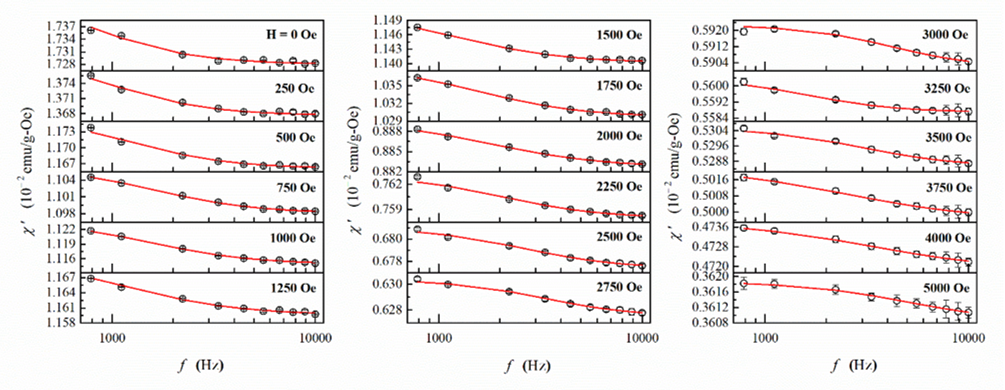}
	\caption{\label{FIG11}(Color online) Frequency dependence of $ \chi^{\prime} $ at various representative magnetic fields (H) in the case of T = 250 K for Mn$ _{1.4} $PtSn. The fits (solid lines) to the data (open symbols) are based on the Eq.(7a) described in the text.}
	\label{Fig11}
\end{figure*}
\begin{figure*}
	\includegraphics[width=14cm,clip]{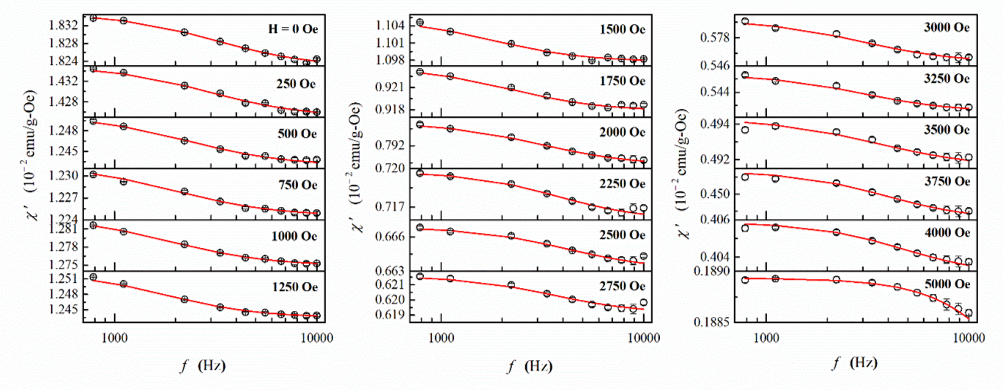}
	\caption{\label{FIG12}(Color online) Frequency dependence of $ \chi^{\prime} $ at various representative magnetic fields (H) in the case of T = 300 K for Mn$ _{1.4} $PtSn. The fits (solid lines) to the data (open symbols) are based on the Eq.(7a) described in the text.}
	\label{Fig12}
\end{figure*}
\begin{figure*}
	\includegraphics[width=14cm,clip]{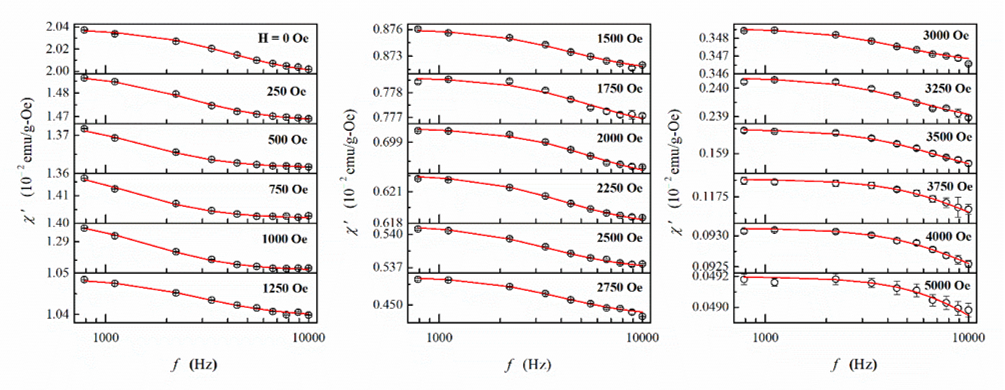}
	\caption{\label{FIG13}(Color online) Frequency dependence of $ \chi^{\prime} $ at various representative magnetic fields (H) in the case of T = 350 K for Mn$ _{1.4} $PtSn. The fits (solid lines) to the data (open symbols) are based on the Eq.(7a) described in the text.}
	\label{Fig13}
\end{figure*}
%%%%%%%%%%%%%%%%%%%%%%%%%%%%%%%%%%%%%%%%%%%%%%%%%%%%%%%%%%%%%%%%%%%%%%%%%%%%%%%%%%%%%%%%%%%%%%%%%%%%%%%%%%%%%%%%%%%%%%%%%%%%%%%%%%%%%%%%%%%%%%%%%%

%%%%%%%%%%%%%%%%%%%%%%%%%%%%%%%%%%%%%%%%%%%%%%%%%%%%%%%%%%%%%%%%%%%%%%%%%%%%%%%%%%%%%%%%%%%%%%%%%%%%%%%%%%%%%%%%%%%%%%%%%%%%%%%%%%%%%%%%%%%%%%%%%%%%%
\begin{figure*}
	\includegraphics[width=14cm,clip]{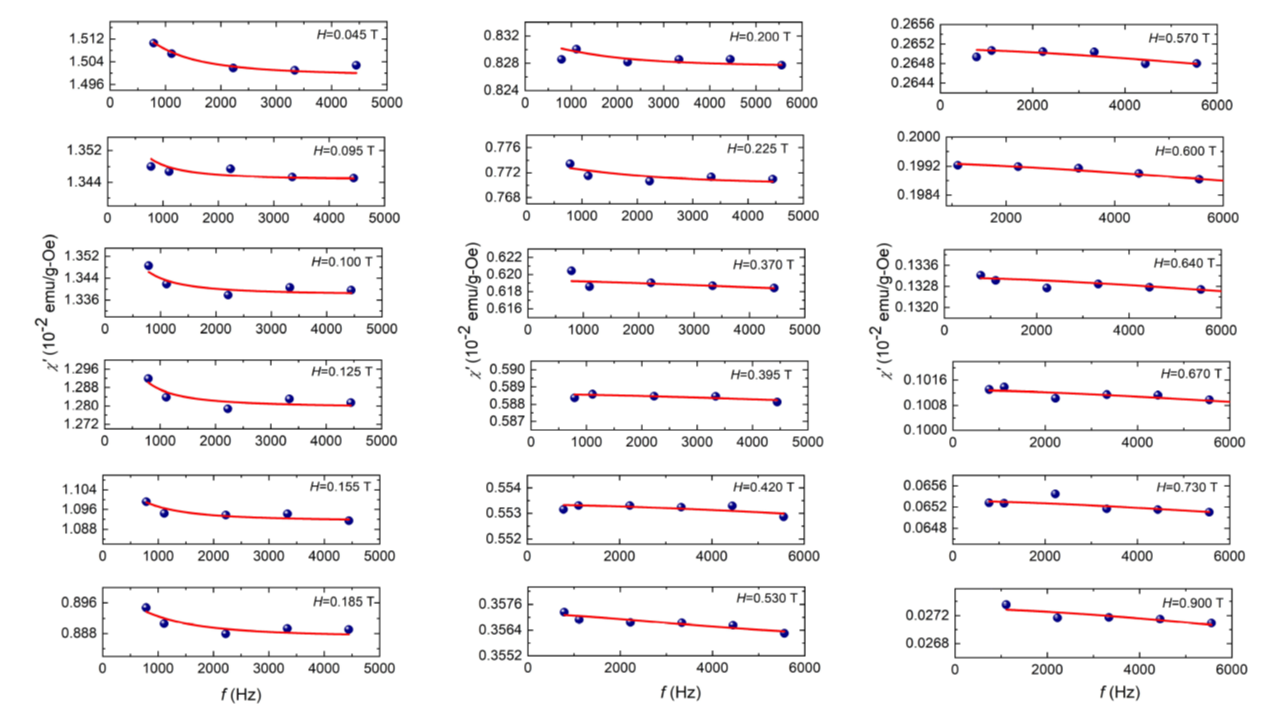}
	\caption{\label{FIG14}(Color online) Frequency dependence of $ \chi^{\prime} $ at various representative magnetic fields (H) in the case of T = 200 K for Mn$ _{1.4} $Pt$ _{0.9} $Pd$ _{0.1} $Sn. The fits (solid lines) to the data (open symbols) are based on the Eq.(7a) described in the text.}
	\label{Fig14}
\end{figure*}

\begin{figure*}
	\includegraphics[width=14cm,clip]{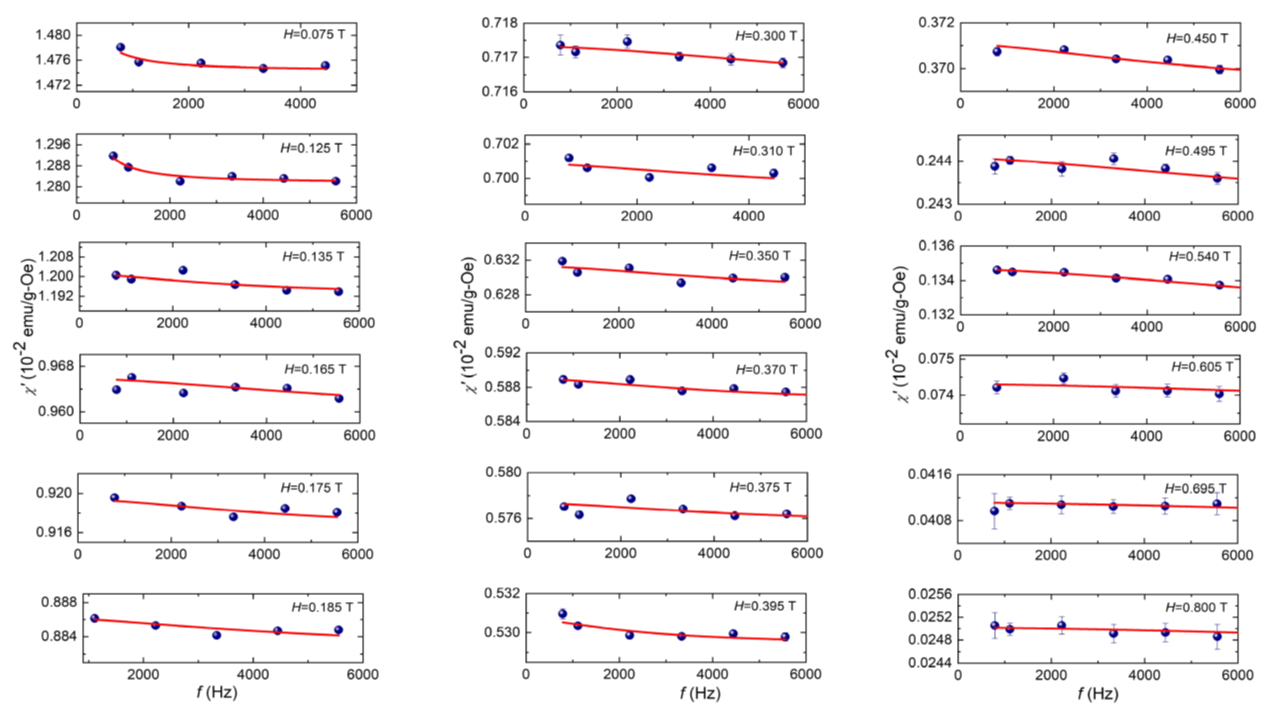}
	\caption{\label{FIG15}(Color online) Frequency dependence of $ \chi^{\prime} $ at various representative magnetic fields (H) in the case of T = 250 K for Mn$ _{1.4} $Pt$ _{0.9} $Pd$ _{0.1} $Sn. The fits (solid lines) to the data (open symbols) are based on the Eq.(7a) described in the text.}
	\label{Fig15}
\end{figure*}

\begin{figure*}
	\includegraphics[width=14cm,clip]{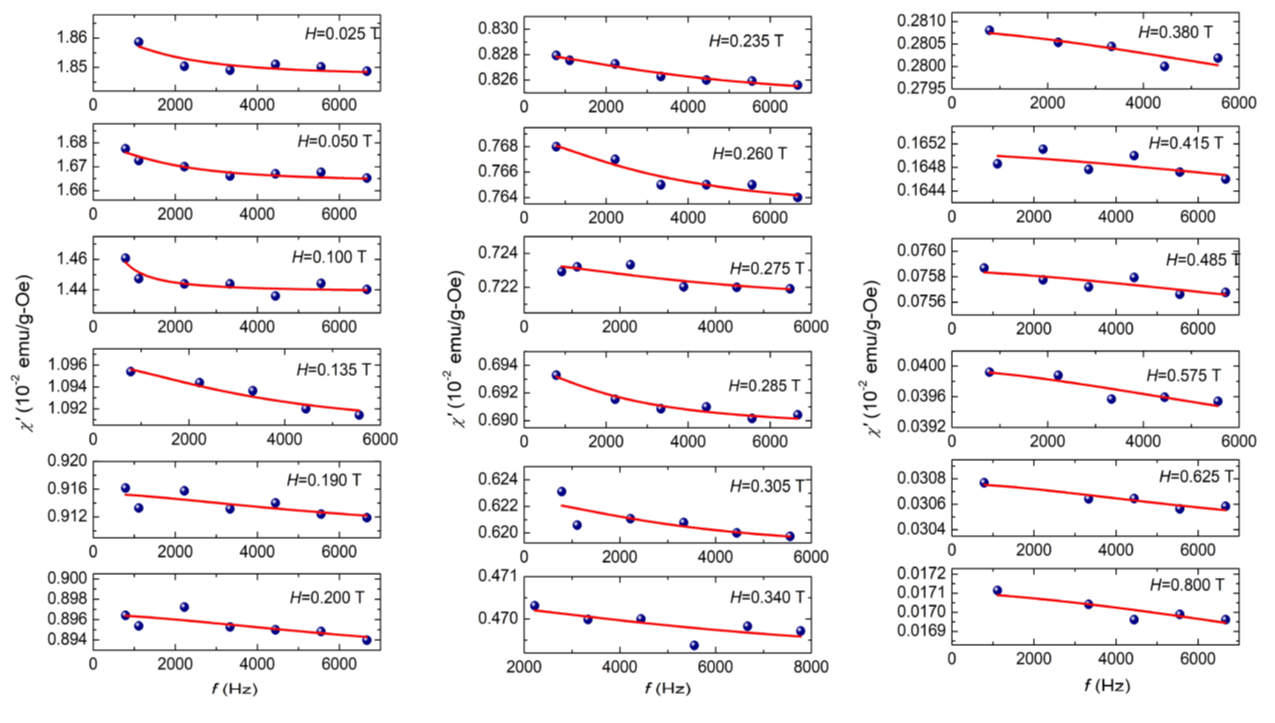}
	\caption{\label{FIG16}(Color online) Frequency dependence of $ \chi^{\prime} $ at various representative magnetic fields (H) in the case of T = 300 K for Mn$ _{1.4} $Pt$ _{0.9} $Pd$ _{0.1} $Sn. The fits (solid lines) to the data (open symbols) are based on the Eq.(7a) described in the text.}
	\label{Fig16}
\end{figure*}

\begin{figure*}
	\includegraphics[width=14cm,clip]{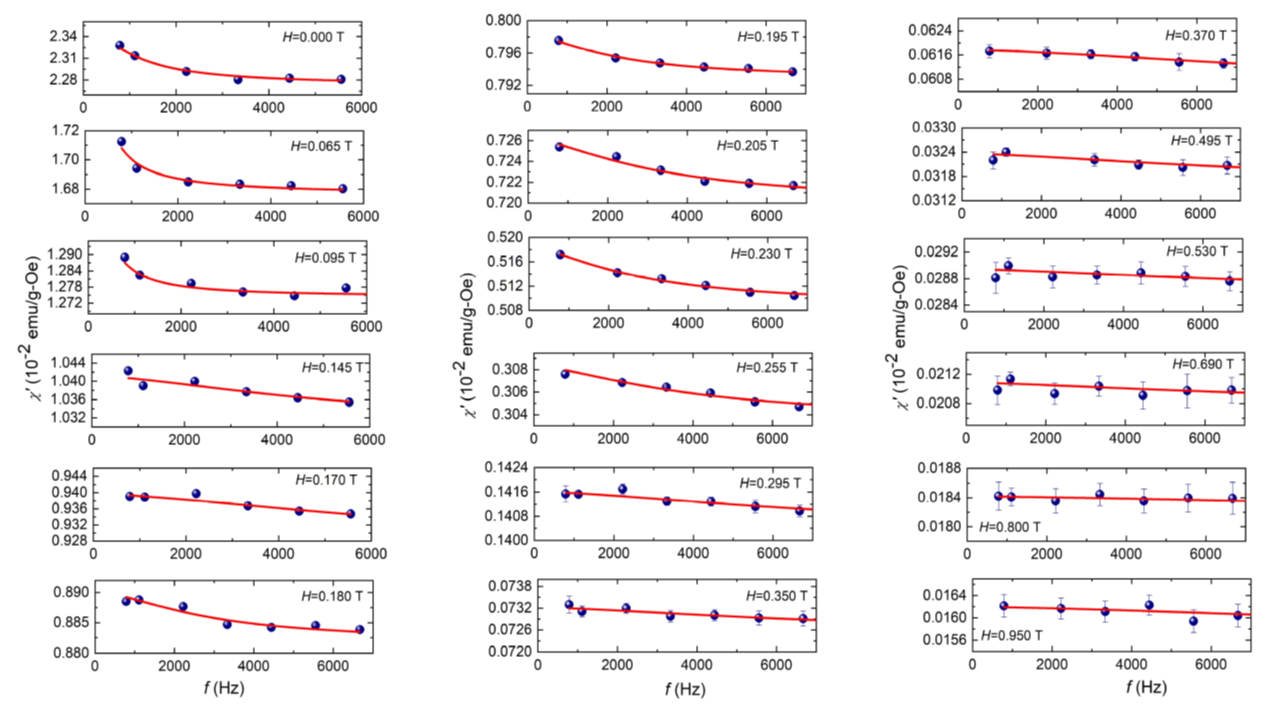}
	\caption{\label{FIG17}(Color online) Frequency dependence of $ \chi^{\prime} $ at various representative magnetic fields (H) in the case of T = 350 K for Mn$ _{1.4} $Pt$ _{0.9} $Pd$ _{0.1} $Sn. The fits (solid lines) to the data (open symbols) are based on the Eq.(7a) described in the text.}
	\label{Fig17}
\end{figure*}
%%%%%%%%%%%%%%%%%%%%%%%%%%%%%%%%%%%%%%%%%%%%%%%%%%%%%%%%%%%%%%%%%%%%%%%%%%%%%%%%%%%%%%%%%%%%%%%%%%%%%%%%%%%%%%%%%%%%%%%%%%%%%%%%%%%%%%%%%%%%%%%%%%

%%%%%%%%%%%%%%%%%%%%%%%%%%%%%%%%%%%%%%%%%%%%%%%%%%%%%%%%%%%%%%%%%%%%%%%%%%%%%%%%%%%%%%%%%%%%%%%%%%%%%%%%%%%%%%%%%%%%%%%%%%%%%%%%%%%%%%%%%%%%%%%%%%%%%
\begin{figure*}
	\includegraphics[width=14cm,clip]{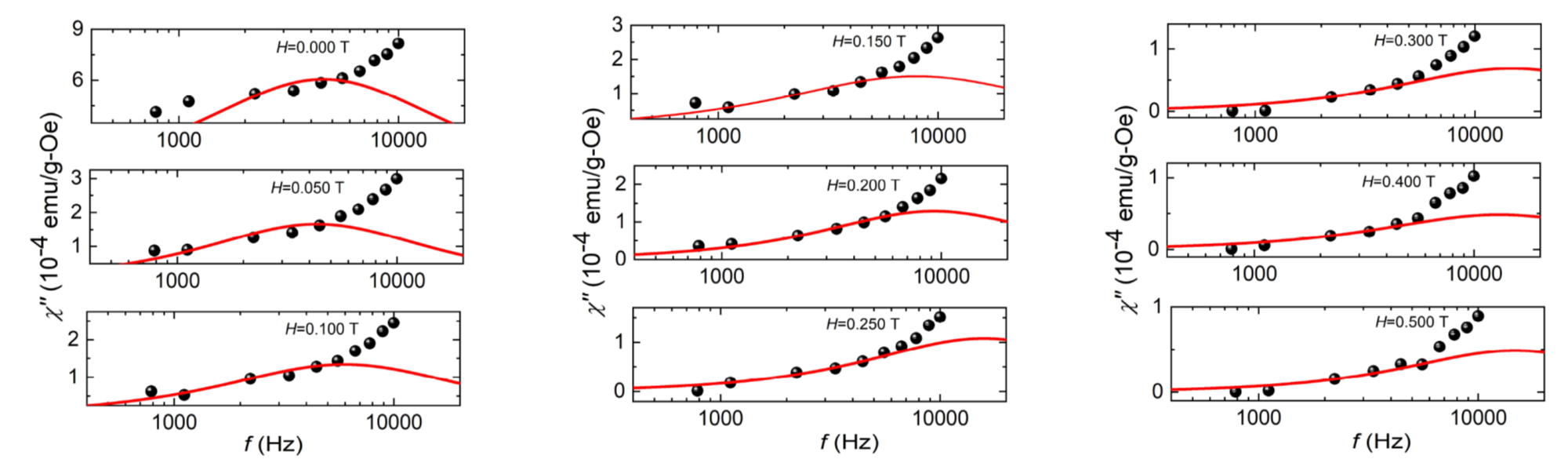}
	\caption{\label{FIG18}(Color online) Frequency dependence of $ \chi^{\prime\prime} $ at various representative magnetic fields (H) in the case of T = 200 K for Mn$ _{1.4} $PtSn. The fits (solid lines) to the data (open symbols) are based on the Eq.(7b) described in the text.}
	\label{Fig18}
\end{figure*}

\begin{figure*}
	\includegraphics[width=14cm,clip]{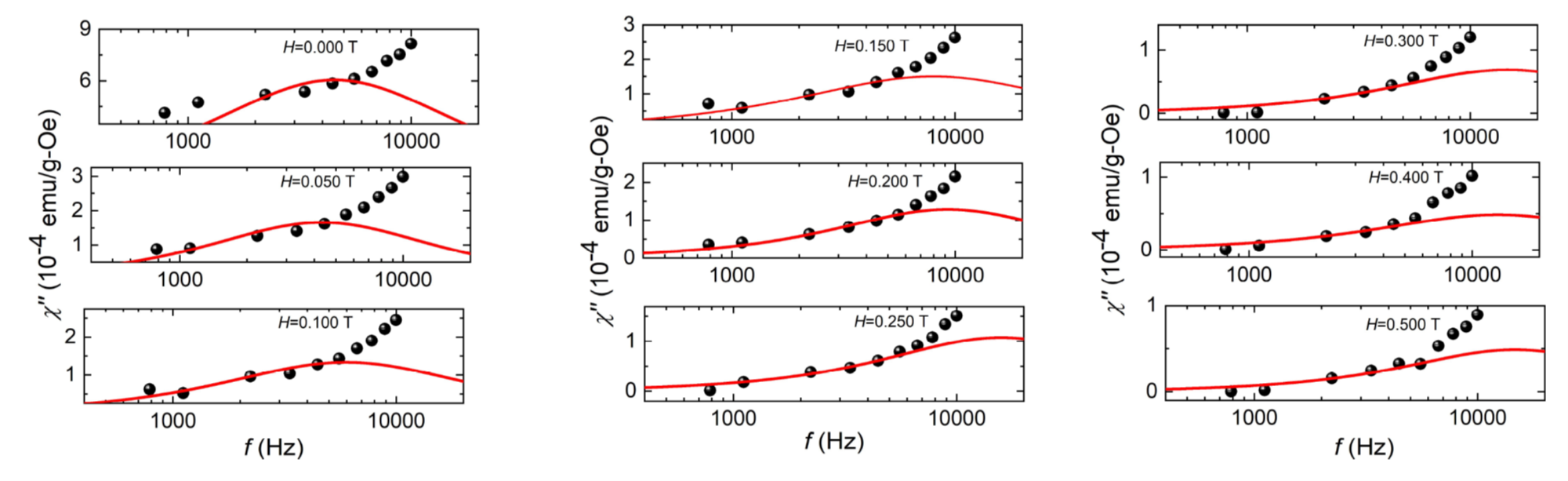}
	\caption{\label{FIG19}(Color online) Frequency dependence of $ \chi^{\prime\prime} $ at various representative magnetic fields (H) in the case of T = 250 K for Mn$ _{1.4} $PtSn. The fits (solid lines) to the data (open symbols) are based on the Eq.(7b) described in the text.}
	\label{Fig19}
\end{figure*}

\begin{figure*}
	\includegraphics[width=14cm,clip]{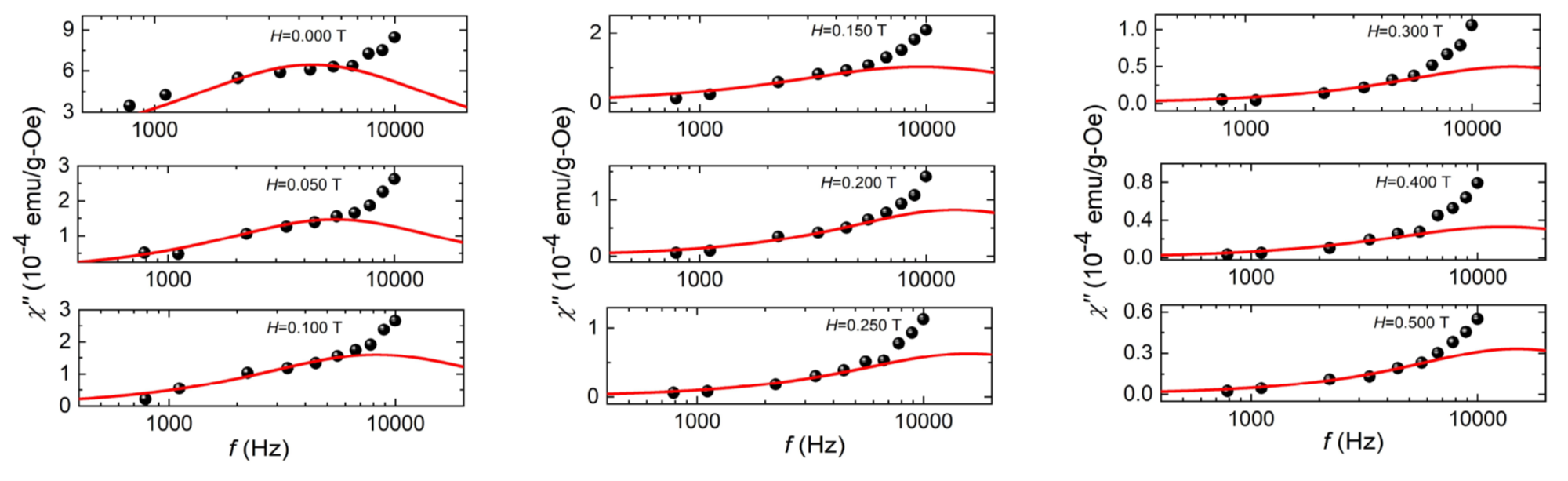}
	\caption{\label{FIG20}(Color online) Frequency dependence of $ \chi^{\prime\prime} $ at various representative magnetic fields (H) in the case of T = 300 K for Mn$ _{1.4} $PtSn. The fits (solid lines) to the data (open symbols) are based on the Eq.(7b) described in the text.}
	\label{Fig20}
\end{figure*}

\begin{figure*}
	\includegraphics[width=14cm,clip]{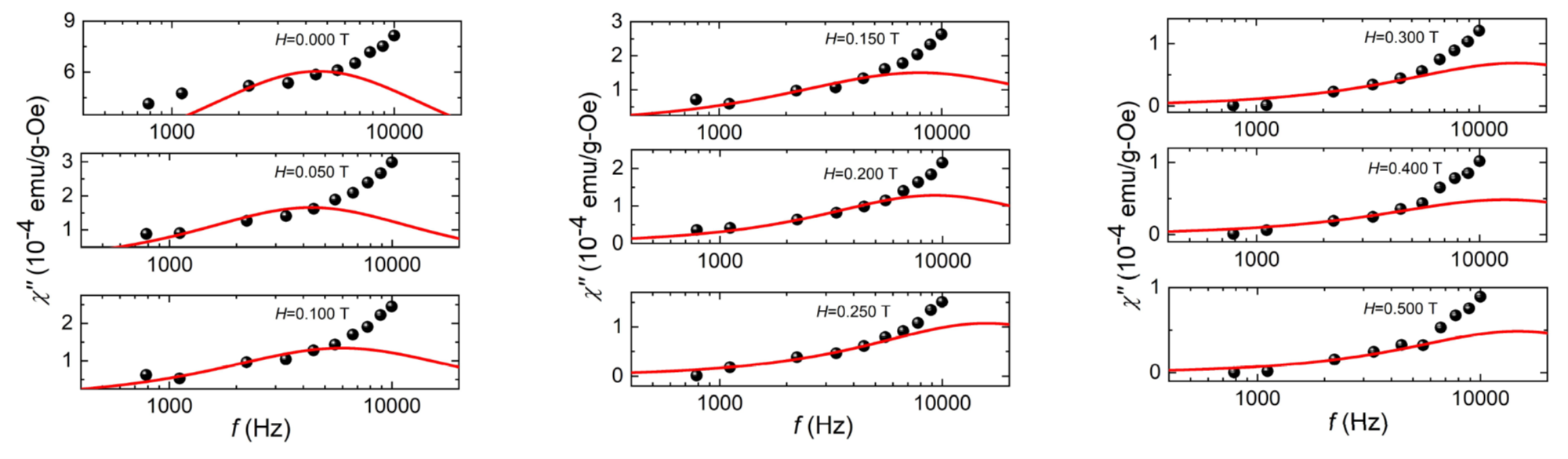}
	\caption{\label{FIG21}(Color online) Frequency dependence of $ \chi^{\prime\prime} $ at various representative magnetic fields (H) in the case of T = 350 K for Mn$ _{1.4} $PtSn. The fits (solid lines) to the data (open symbols) are based on the Eq.(7b) described in the text.}
	\label{Fig21}
\end{figure*}
%%%%%%%%%%%%%%%%%%%%%%%%%%%%%%%%%%%%%%%%%%%%%%%%%%%%%%%%%%%%%%%%%%%%%%%%%%%%%%%%%%%%%%%%%%%%%%%%%%%%%%%%%%%%%%%%%%%%%%%%%%%%%%%%%%%%%%%%%%%%%%%%%%

%%%%%%%%%%%%%%%%%%%%%%%%%%%%%%%%%%%%%%%%%%%%%%%%%%%%%%%%%%%%%%%%%%%%%%%%%%%%%%%%%%%%%%%%%%%%%%%%%%%%%%%%%%%%%%%%%%%%%%%%%%%%%%%%%%%%%%%%%%%%%%%%%%%%%
\begin{figure*}
	\includegraphics[width=14cm,clip]{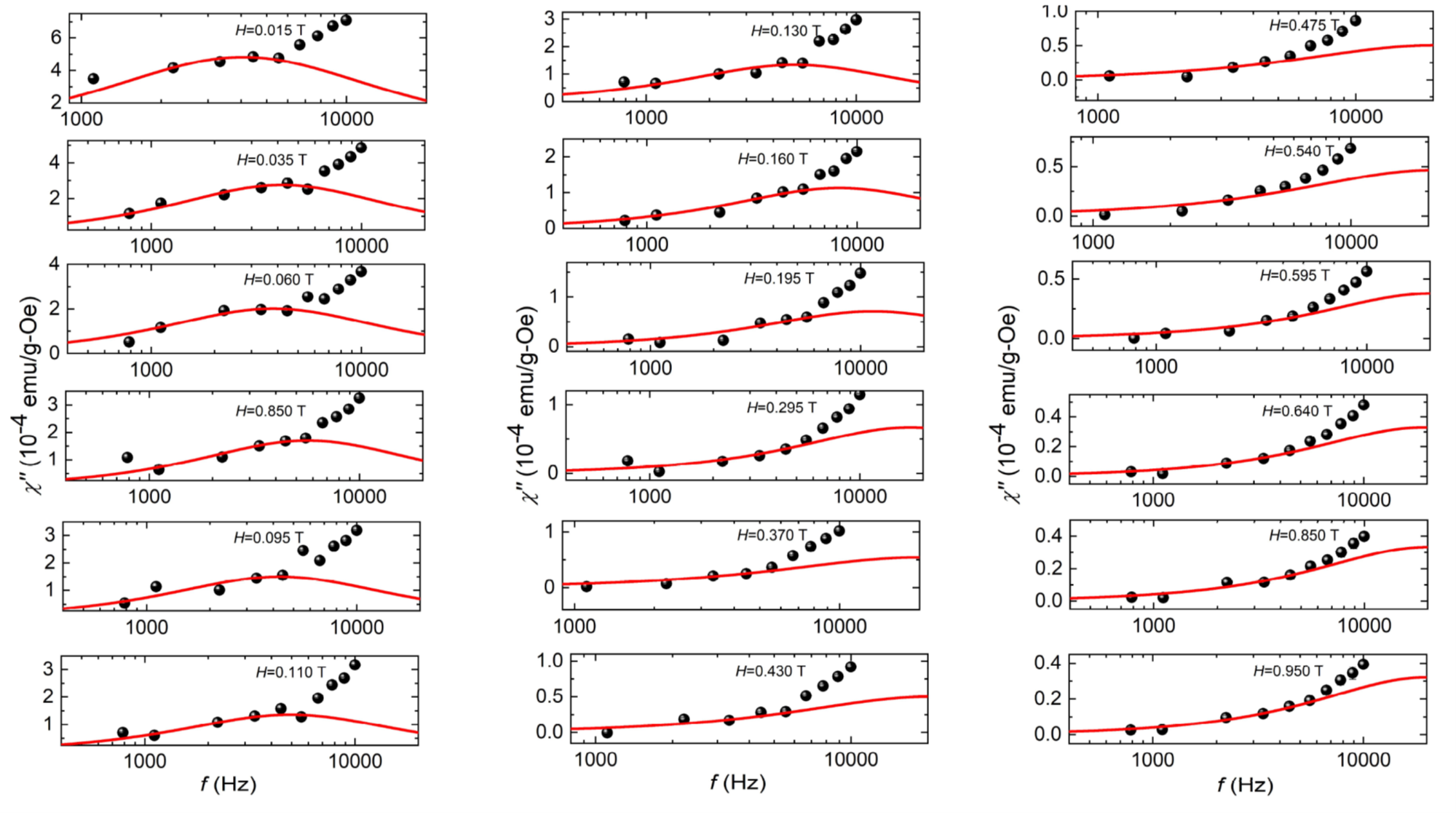}
	\caption{\label{FIG22}(Color online) Frequency dependence of $ \chi^{\prime\prime} $ at various representative magnetic fields (H) in the case of T = 200 K for Mn$ _{1.4} $Pt$ _{0.9} $Pd$ _{0.1} $Sn. The fits (solid lines) to the data (open symbols) are based on the Eq.(7b) described in the text.}
	\label{Fig22}
\end{figure*}

\begin{figure*}
	\includegraphics[width=14cm,clip]{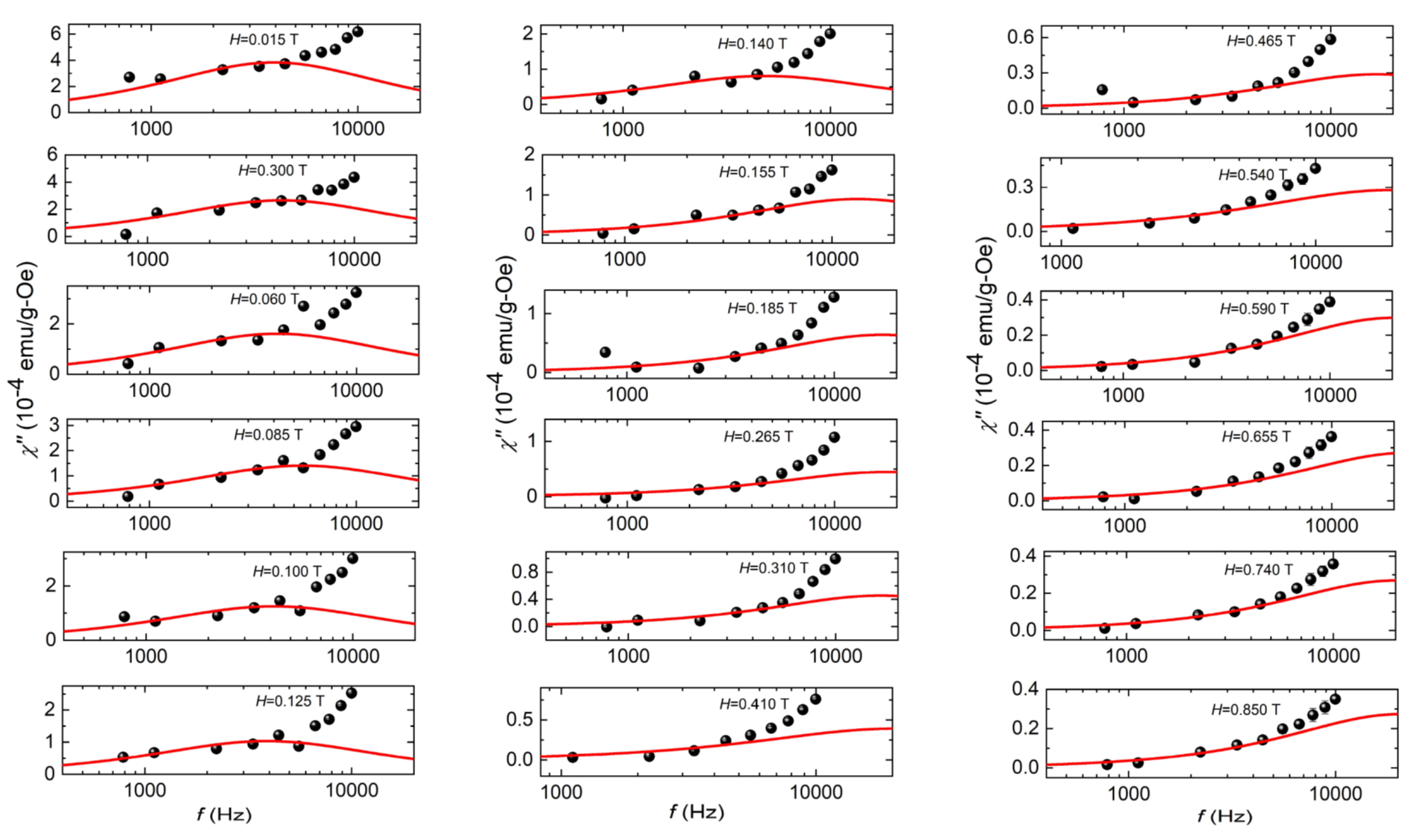}
	\caption{\label{FIG23}(Color online) Frequency dependence of $ \chi^{\prime\prime} $ at various representative magnetic fields (H) in the case of T = 250 K for Mn$ _{1.4} $Pt$ _{0.9} $Pd$ _{0.1} $Sn. The fits (solid lines) to the data (open symbols) are based on the Eq.(7b) described in the text.}
	\label{Fig23}
\end{figure*}

\begin{figure*}
	\includegraphics[width=14cm,clip]{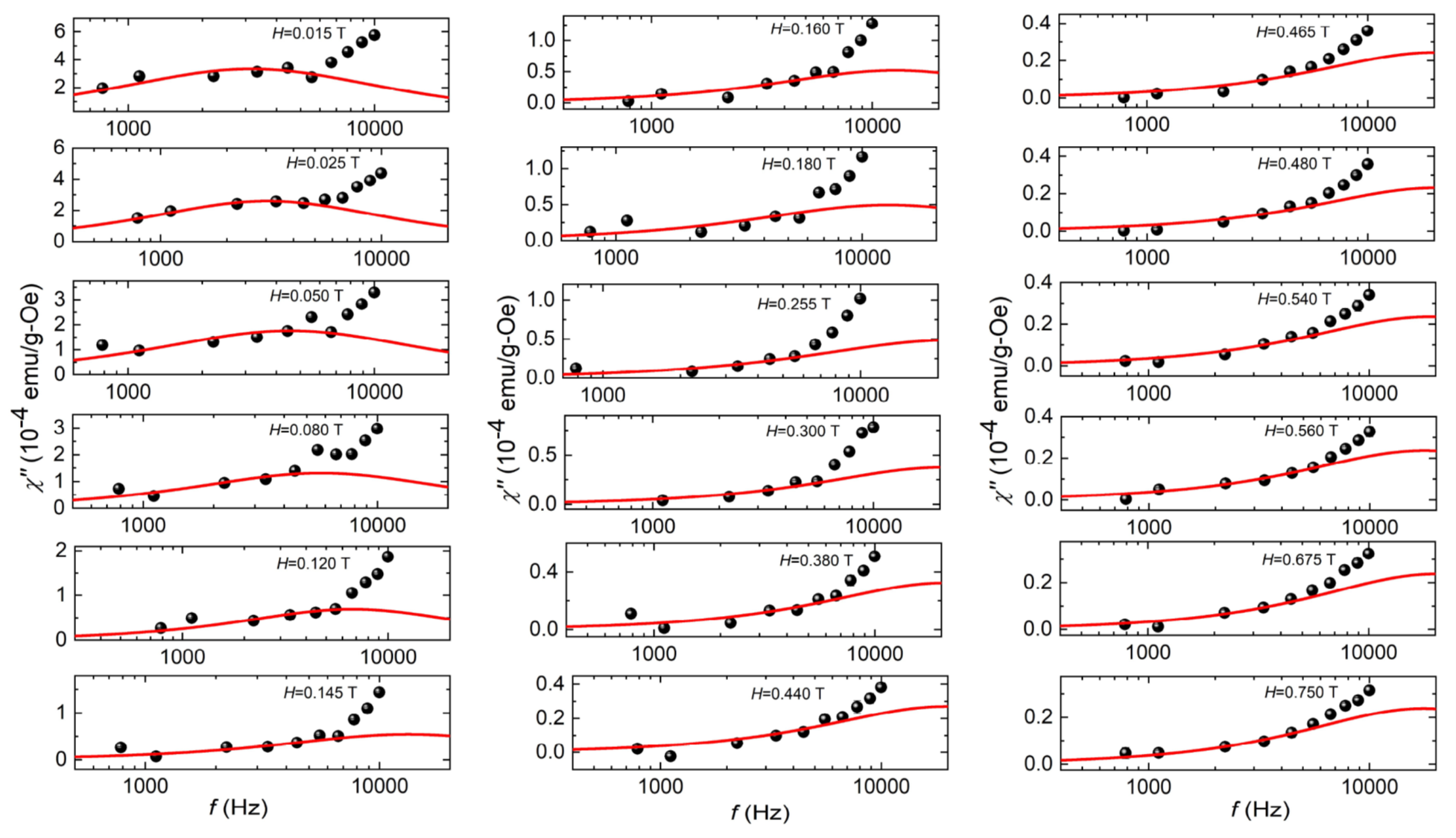}
	\caption{\label{FIG24}(Color online) Frequency dependence of $ \chi^{\prime\prime} $ at various representative magnetic fields (H) in the case of T = 300 K for Mn$ _{1.4} $Pt$ _{0.9} $Pd$ _{0.1} $Sn. The fits (solid lines) to the data (open symbols) are based on the Eq.(7b) described in the text.}
	\label{Fig24}
\end{figure*}

\begin{figure*}
	\includegraphics[width=14cm,clip]{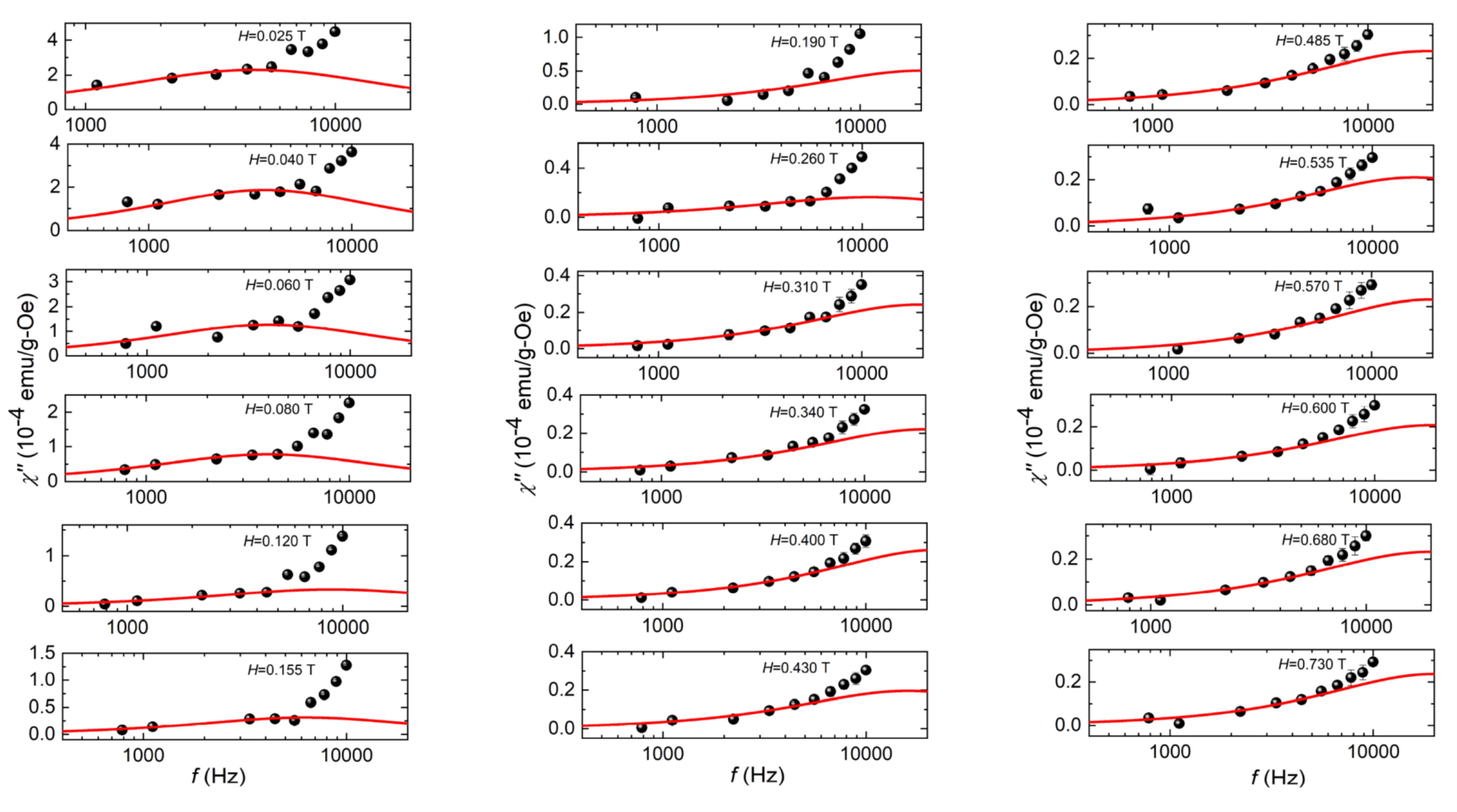}
	\caption{\label{FIG25}(Color online) Frequency dependence of $ \chi^{\prime\prime} $ at various representative magnetic fields (H) in the case of T = 350 K for Mn$ _{1.4} $Pt$ _{0.9} $Pd$ _{0.1} $Sn. The fits (solid lines) to the data (open symbols) are based on the Eq.(7b) described in the text.}
	\label{Fig25}
\end{figure*}
%%%%%%%%%%%%%%%%%%%%%%%%%%%%%%%%%%%%%%%%%%%%%%%%%%%%%%%%%%%%%%%%%%%%%%%%%%%%%%%%%%%%%%%%%%%%%%%%%%%%%%%%%%%%%%%%%%%%%%%%%%%%%%%%%%%%%%%%%%%%%%%%%%

\end{document}